\documentclass[10pt, twocolumn, twoside]{IEEEtran}

\usepackage[Symbol]{upgreek}

\usepackage{amsthm,amsmath,amssymb,amsfonts,mathrsfs,mathtools,dsfont,mathrsfs,bm,bbm}
\DeclareMathAlphabet{\mathpzc}{OT1}{pzc}{m}{it}
\DeclareSymbolFontAlphabet{\amsmathbb}{AMSb}%
\usepackage{csvsimple,booktabs}
\usepackage{float,enumitem}
\usepackage[caption=false]{subfig}
\usepackage[dvipsnames]{xcolor}
\usepackage{graphicx,epsfig}
\usepackage{balance}
\usepackage[hyphens]{url}
\usepackage[breaklinks, colorlinks, linkcolor=MidnightBlue, anchorcolor=MidnightBlue, citecolor=MidnightBlue, urlcolor=MidnightBlue]{hyperref}
\usepackage{cite}

\usepackage[ruled,linesnumbered]{algorithm2e}

\newcommand{\Eset}{\mathbb{E}}

\newcommand{\Rset}{\mathbb{R}}

\newcommand{\Bcal}{{\cal B}}

\newcommand{\Fcal}{{\cal F}}

\newcommand{\Hcal}{{\cal H}}

\newcommand{\Ocal}{{\cal O}}
\newcommand{\Pcal}{{\cal P}}

\newcommand{\Scal}{{\cal S}}
\newcommand{\Tcal}{{\cal T}}

\newcommand{\Xcal}{{\cal X}}

\newcommand{\xbar}{{\bar{x}}}

\newtheorem{thm}{Theorem}

\newtheorem{definition}{Definition}

\newtheorem{assump}{Assumption}
\newtheorem{remark}{Remark}

\begin{document}

\title{\LARGE {\bf{Byzantine Fault-Tolerance in Federated Local SGD\\ under $2f$-Redundancy}}}

\author{Nirupam Gupta${^1}$ \hspace{0.5in} Thinh T. Doan${^2}$ \hspace{0.5in} Nitin Vaidya${^3}$ \\
% \vspace{0.5cm}
% ${^1}$EPFL, Switzerland \hspace{0.5in} ${^2}$Virginia Tech., USA \hspace{0.5in} ${^3}$Georgetown University, USA 
	
% \thanks{$^*$The work was partially supported by the ARL under Cooperative Agreement W911NF-17-2-0196, and by the NSF award 1842198.}
\thanks{${^1}$EPFL, Switzerland (\texttt{nirupam115@gmail.com}). ${^2}$Virginia Tech., USA (\texttt{thinhdoan@vt.edu}). ${^3}$Georgetown University, USA (\texttt{nitin.vaidya@georgetown.edu}).}
}
% make the title area
\maketitle

\begin{abstract}

We consider the problem of Byzantine fault-tolerance in federated machine learning. In this problem, the system comprises multiple agents each with local data, and a trusted centralized coordinator. In fault-free setting, the agents collaborate with the coordinator to find a minimizer of the aggregate of their local cost functions defined over their local data. We consider a scenario where some agents ($f$ out of $N$) are Byzantine faulty. Such agents need not follow a prescribed algorithm correctly, and may communicate arbitrary incorrect information to the coordinator. In the presence of Byzantine agents, a more reasonable goal for the non-faulty agents is to find a minimizer of the aggregate cost function of only the non-faulty agents. This particular goal is commonly referred as {\em exact fault-tolerance}. Recent work has shown that exact fault-tolerance is achievable if only if the non-faulty agents satisfy the property of {\em $2f$-redundancy}. Now, under this property, techniques are known to impart exact fault-tolerance to the distributed implementation of the classical stochastic gradient-descent (SGD) algorithm. However, we do not know of any such techniques for the federated local SGD algorithm - a more commonly used method for federated machine learning. To address this issue,
% Under this property, the non-faulty agents can achieve exact fault-tolerance using the distributed implementation of the classical gradient-descent (GD) algorithm equipped with {\em comparative gradient clipping} (CGC). 
we propose a novel technique named {\em comparative elimination} (CE). We show that, under $2f$-redundancy, the federated local SGD algorithm with CE can indeed obtain exact fault-tolerance in the deterministic setting when the non-faulty agents can accurately compute gradients of their local cost functions. In the general stochastic case, when agents can only compute unbiased noisy estimates of their local gradients, our algorithm achieves {\em approximate} fault-tolerance with approximation error proportional to the variance of stochastic gradients and the fraction of Byzantine agents.

% formal fault-tolerance properties of our algorithm in both the deterministic and stochastic settings, under $2f$-redundancy. Specifically, in the deterministic case, our algorithm achieves exact fault-tolerance against a bounded fraction of Byzantine agents. 
% exists {\em gradient-filtering} techniques that solves this problem when the non-faulty agents implement the distributed gradient-descent 
% For achieving this goal, we propose a {\em local gradient-descent} (GD) algorithm incorporating a novel {\em comparative elimination} (CE) filter (aka.~aggregation scheme) to provably mitigate the detrimental impact of Byzantine agents. 
% In the deterministic setting, we show that our algorithm can guarantee {\em exact} fault-tolerance against a bounded fraction of Byzantine agents, provided the non-faulty agents satisfy the known necessary condition of {\em $2f$-redundancy}. In the stochastic setting, when the agents can only compute stochastic estimates of their gradients, our algorithm guarantees {\em approximate} fault-tolerance where the approximation error is proportional to the variance of stochastic gradients and the fraction of Byzantine agents.
% that communicate with a centralized coordinator. We allow up to $f$ Byzantine-faulty agents, which may not follow a prescribed algorithm correctly, and may share arbitrary incorrect information with the coordinator. Associated with each non-faulty agent is a local cost function. The goal of the non-faulty agents is to compute a minimizer of their aggregate cost function. 

\end{abstract}

\begin{IEEEkeywords}
Federated optimization, Byzantine fault-tolerance, local gradient-descent
\end{IEEEkeywords}

%--------------------------

%!TEX root = main.tex

\section{Introduction}
We consider a distributed optimization framework where there are $N$ agents communicating with a single coordinator.  Associated with each agent $i$ is a function $q^{i}:\mathbb{R}^{d}\rightarrow \mathbb{R}$. The goal of the agents is to find $x^{\star}$ such that
% solves\vspace{-0.1cm}
\begin{align}
x^{\star} \in \underset{x \in \mathbb{R}^d}{\arg \min} \sum_{i=1}^{N}q^{i}(x) ,\label{prob:obj}    
% \vspace{-0.1cm}
\end{align}
where each $q^{i}$ is given as 
\begin{align}
q^{i}(x) \triangleq \Eset_{\pi^{i}}\left[Q^{i}(x;X^{i})\right],\label{prob:Fi}     
\end{align}
for some random variable $X^{i}$ defined over a sample set $\Xcal^{i}$ with a distribution $\pi^{i}$. We assume that each agent $i$ only has access to a sequence of vectors $\{G^{i}(\cdot)\}$, which can be either the {\em actual} gradients $\{\nabla q^{i}(\cdot)\}$ or {\em stochastic} gradients $\{\nabla Q^{i}(\cdot,X^{i})\}$ of its cost function $q^i(\cdot)$. This is a common distributed machine learning setting when there is a large number of data distributed to different agents (or machines). The goal is to design an algorithm that allows these agents is to jointly minimize a loss function defined over their data, i.e., solve the optimization problem~\eqref{prob:obj}.     

For solving this problem, we consider the {\em federated local stochastic gradient-descent} (local SGD) method, which has recently received significant attention due to its application in distributed learning \cite{Kairouz_survery2020,Li_survery2020}. In this method, the coordinator maintains an estimate of a solution defined in~\eqref{prob:obj}. This estimate is broadcast to all the agents, and each agent updates its copy of the estimate by running a number of local SGD steps. The agents send back to the coordinator their local updated estimates. Finally, the coordinator averages the received estimates to obtain new {\em global} estimate of $x^{\star}$. Eventually, if all the agents are non-faulty, the sequence of global estimates converges to a solution~\eqref{prob:obj}. Since the agents only share their local estimates and not their data, federated local SGD is widely used in privacy conscious distributed learning~\cite{Kairouz_survery2020}.   

% each agent runs a number of GD steps locally based on its local data to update their local variables, which are then aggregated at the coordinator to find $x^{\star}$, a solution of \eqref{prob:obj}. 

Our interest is to study the federated local SGD algorithm when up to $f$ agents are Byzantine faulty~\cite{lamport1982byzantine}. Such faulty agents may behave arbitrarily, and their identity is a priori unknown. In particular, Byzantine faulty agents may collude and share incorrect information with the coordinator in order to corrupt the output of the algorithm, e.g., see~\cite{xie2019dba}. We aim to design a new local SGD method that allows all the non-faulty agents to compute an exact minimum of the aggregate cost of the non-faulty agents, despite the presence of Byzantine agents. Specifically, we consider the problem of {\em exact fault-tolerance} defined below. For a set $\Hcal$, $|\Hcal|$ denotes its cardinality. 
\begin{definition}[{\bf Exact fault-tolerance}]
\label{def:eft}
\!Let $\Hcal$ with $|\Hcal| \geq N-f$ be the set of non-faulty agents. A distributed optimization algorithm is said to have exact fault-tolerance if it allows all the non-faulty agents to compute
% $x^{\star}_{\Hcal}$ such that
\begin{align}
    x^{\star}_{\Hcal} \in \underset{x \in \mathbb{R}^d}{\arg \min} \sum_{i \in \Hcal} q^{i}(x). \label{prob:hon_obj}
    % \vspace{-0.2cm}
\end{align}
\end{definition}
Since the identity of the Byzantine faulty agents is a priori unknown, in general, {exact fault-tolerance} is unachievable~\cite{su2016fault}. Indeed, recent work has shown that exact fault-tolerance can be achieved if and only if the non-faulty agents satisfy the property of {\em $2f$-redundancy} defined as follows~\cite{gupta2020fault_podc, gupta2020resilience}.
\begin{definition}[{ $2f$-\bf{redundancy}}]
\label{def:2t_red}
A set of non-faulty agents $\Hcal$, with $|\Hcal| \geq n-f$, is said to have $2f$-redundancy if for any subset $\Scal \subseteq \Hcal$ with $|\Scal| \geq N-2f$, 
% the following condition holds
\begin{equation}
\underset{x \in \mathbb{R}^d}{\arg \min} \, \sum_{i \in \Scal} q_i(x) = \underset{x \in \mathbb{R}^d}{\arg \min} \, \sum_{i \in \Hcal} q_i(x).
% \vspace{-0.2cm}
\end{equation}
\end{definition}

The $2f$-redundancy property is critical to our algorithm, presented in Section~\ref{sec:alg}, for solving~\eqref{prob:hon_obj}. 
This property implies that a minimizer of the aggregate cost of any $N-2f$ non-faulty agents is also a minimizer of the aggregate cost of all the non-faulty agents, and vice-versa. This seemingly contrived condition arises naturally with high probability in many practical applications of the distributed optimization problem, such as distributed hypothesis testing~\cite{chong2015observability, gupta2019byzantine, mishra2016secure, su2019finite}, and distributed learning~\cite{alistarh2018byzantine, blanchard2017machine, charikar2017learning, guerraoui2018hidden}. In the context of distributed learning, when all agents have the same data generating distribution (i.e., the homogeneous data setting), $2f$-redundancy hold true trivially. In general, however, $2f$-redundancy also holds true in the non-homogeneous data setting as long as we can solve the non-faulty distributed learning problem~\eqref{prob:hon_obj} using information (or data) from only $N-2f$ non-faulty agents. More details on $2f$-redundancy, including a formal proof for its necessity, can be found in~\cite{gupta2020fault_podc, gupta2020resilience,liu2021approximate}.

Prior work indicates that solving the exact fault-tolerance problem~\eqref{prob:hon_obj} is nontrivial even under the $2f$-redundancy property, especially in the high dimension case, e.g., see~\cite{gupta2020fault_podc, kuwaranancharoen2020byzantine, su2019finite}. This can be attributed to two main factors; (i) the identity of Byzantine faulty agents is a priori unknown, and (ii) smart Byzantine agents can inject malicious information without getting detected (e.g., see~\cite{xie2020fall}). Although there exist techniques that impart {\em provable} exact fault-tolerance to the distributed implementation of SGD algorithm (in which the agents simply share their local gradients, and do not maintain local estimates)~\cite{blanchard2017machine, gupta2020fault_podc, yin2018byzantine}, we do not know of such techniques for the federated local SGD algorithm. This motivates us to propose a technique named {\em comparative elimination} (CE) that provably robustifies the federated local SGD algorithm against Byzantine agents. In the special deterministic setting, we show that CE can provide exact fault-tolerance. In the generic stochastic setting, we achieve {\em approximate} fault-tolerance with approximation error proportional to the variance of stochastic gradients and fraction of Byzantine agents $f/N$.

Intuitively speaking, the CE techniques allows the coordinator to mitigate the detrimental impact of potentially adversarial estimates sent by the Byzantine agents. Specifically, instead of simply averaging the agents' local estimates, the coordinator eliminates $f$ estimates that are farthest from the current global estimate maintained by the coordinator. Then, the remaining $N-f$ agents' estimates are averages to obtain the new global estimate. Details of our scheme, along with its formal fault-tolerance properties, is presented in Section \ref{sec:alg}. 

We present below a summary of our main contributions, and then discuss the related literature.
% In a sense, since the coordinator does not know the identities of Byzantine models it tries to remove at most $2f$ values that have large distances to its current iterate. 

\subsection{\bfseries Main Contributions}
% We study the problem of {\em fault-tolerance} problem, stated in Definition~\ref{def:eft}, in the context of local gradient-descent (GD) method for distributed optimization. 
We formally analyse the Byzantine robustness of the federated local SGD algorithm when coupled with the aforementioned technique of {\em comparative elimination} (CE). Specifically, assuming each non-faulty agent's cost function to be {\em $L$-smooth}, the aggregate non-faulty cost to be {\em $\mu$-strongly convex} (but, the local costs may only be convex), and the necessary condition of {\em $2f$-redundancy}, we present the following results:

\begin{itemize}[leftmargin = 5mm]
    \item {\bf In the deterministic case}, when each non-faulty agent $i$ updates its local estimates using {\em actual} gradients of its cost $\{\nabla q^{i}(\cdot)\}$, the CE filter scheme guarantees {\em exact fault-tolerance} if $\frac{f}{|H\cal|} \leq \frac{\mu}{3 L}$. Moreover, the convergence is linear, similar to the fault-free setting under strong convexity.
    \item {\bf In the stochastic case}, when a non-faulty agent $i$ can only compute unbiased noisy estimates of its local gradients $\{\nabla Q^{i}(\cdot,X^{i})\}$ we guarantee {\em approximate} fault-tolerance. Specifically, the sequence of global estimates $\{\xbar^{k}\}$ satisfy the following for all $k$:
    \begin{align*}
    \Eset[\|\xbar^{k} - x_{\Hcal}^{\star}\|^2] \leq \lambda^{k}\Eset[\|\xbar_{0} - x_{\Hcal}^{\star}\|^2] + \Ocal \left(\sigma^2 \alpha + \frac{\sigma^2 f}{N-f} \right)
    \end{align*}
    for some $\lambda \in (0, 1)$ where $\alpha$ is a constant step-size of the algorithm, and $\sigma$ is the variance of stochastic gradients. 
\end{itemize} 

Specific details of our results are given in Section \ref{sec:results}.   
% We show that if the local objective functions have $2f$-redundancy property (Definition \ref{def:2t_red}) then the proposed algorithm provably mitigates the impact of Byzantine agents, and solve problem~\eqref{prob:hon_obj} efficiently. In particular, when the global objective function is $L$-smooth and $\mu$-strongly convex (but, the local functions are only convex), and the $2f$-redundancy is satisfied with 
% \begin{align*}
% \frac{f}{|H\cal|} \leq \frac{\mu}{3 L},     
% \end{align*}
% our algorithm converges to $x_{\Hcal}^{\star}$ at a linear rate in the deterministic setting (i.e., when $\{\nabla q^{i}(\cdot)\}$ is known)
% \begin{align*}
% \|\xbar^{k} - \xbar_{\Hcal}^{\star}\|^2 \leq \lambda^{k}\left\|\xbar_{0} - x_{\Hcal}^{\star}\right\|^2,
% \end{align*}
% for some $\lambda \in (0,1)$.  

\subsection{Related Work}
\label{sub:rel}

In recent years, several schemes have been proposed for Byzantine fault-tolerance in distributed implementation of SGD algorithm. Prominent schemes include {\em coordinate-wise trimmed mean} (CWTM)~\cite{su2016fault, su2019finite,yang2019byrdie,yin2018byzantine}, {\em multi-KRUM}~\cite{blanchard2017machine}, {\em geometric median-of-means} (GMoM)~\cite{chen2017distributed}, {\em coordinate-wise median}~\cite{sundaram2018distributed, yin2018byzantine},  {\em Bulyan}~\cite{guerraoui2018hidden}, {\em minimum-diameter averaging} (MDA)~\cite{guerraoui2018hidden}, {\em phocas}~\cite{xie2018phocas}, {\em Byzantine-robust stochastic aggregation} (RSA)~\cite{li2019rsa}, {\em signSGD} with majority voting~\cite{sohn2020election}, and {\em spectral decomposition} based filters~\cite{diakonikolas2019sever, prasad2020robust}. Most of these works, with the exception of~\cite{su2016fault, su2019finite, sundaram2018distributed, kuwaranancharoen2020byzantine, li2019rsa, yang2019byrdie}, only consider the framework of distributed SGD wherein the agents send gradients of their local costs, and their results are not readily applicable to the federated local SGD framework that we consider. Nevertheless, these works suggest that the aforementioned schemes need not guarantee {\em exact fault-tolerance} in general, even in the deterministic setting with $2f$-redundancy, unless further assumptions are made on the non-faulty agents' costs. 

Although~\cite{su2016fault, su2020byzantine, sundaram2018distributed} implicitly show exact fault-tolerance properties of {\em trimmed-mean} and {\em median}, they only consider the scalar case where agents' cost functions are univariate. The extension of their results to higher-dimensions is non-trivial and remains poorly understood, e.g., see~\cite{kuwaranancharoen2020byzantine, su2016fault, su2019finite, yang2017byrdie}. For instance,~\cite{su2016fault} considers a degenerate case wherein the agents' cost functions have a known common {\em basis}. The prior work~\cite{su2019finite} shows that CWTM can guarantee exact fault-tolerance when solving the distributed linear least squares problem provided the agents' data satisfies a condition stronger than $2f$-redundancy. In~\cite{yang2019byrdie}, they assume that the agents' costs can be decomposed into independent {\em scalar strictly convex} functions. Recently,~\cite{kuwaranancharoen2020byzantine} studied the fault-tolerance of CWTM in a peer-to-peer setting (a generalization of federated model) for generic convex optimization problems; their results suggest that CWTM need not provide exact fault-tolerance even under $2f$-redundancy.

% The problem of distributed optimization finds direct application in distributed state estimation~\cite{rabbat2004distributed}. In this problem, the system comprises multiple sensors, and each sensor makes partial observations about the system's state. The goal is to compute the entire state of the system using collective observations from all the sensors. However, if a sensor is faulty then it may share incorrect observations, preventing correct state estimation. The special case of distributed state estimation when the observations are {\em linear} in the system's state has gained significant attention in the past, e.g. see~\cite{bhatia2015robust, chong2015observability, mishra2016secure, pajic2017attack, pajic2014robustness, shoukry2015imhotep, shoukry2017secure, su2019finite}. These works have shown that the state can be determined despite up to $f$ (out of $n$) faulty observations {\em if and only if} the system is {\em $2f$-sparse observable}, i.e., the complete state can be determined using observations of only $n-2f$ non-faulty sensors. We note that, in this particular case, {\em $2f$-sparse observability} is equivalent to $2f$-redundancy. Additionally, some of these works, such as~\cite{mishra2016secure, su2019finite}, also consider the case of {\em approximate} linear state estimation when the observations are noisy. Our work is more general in that we consider the problem setting of distributed optimization, and our results apply to a larger class of cost functions.

When applied to the federated local SGD framework, some of the above schemes, including multi-KRUM, Bulyan, CWTM, GMoM and MDA, operate only on the local estimates sent by the agents and disregard the current global estimate maintained by the coordinator~\cite{fang2020local}. On the other hand, our proposed CE filter exploits the (supposed) closeness between the current global estimate and the non-faulty agents' local updated estimates to improve robustness against Byzantine agents. The closeness between the global and non-faulty agents' local estimates exists due to Lipschitz smoothness of agents' local cost functions. This is a critical observation for the fault-tolerance property of our algorithm. Other works that also exploit this observation are RSA~\cite{li2019rsa}, and~\cite{munoz2019byzantine}.

Recently,~\cite{wu2020federated} have shown that {\em geometric median aggregation} scheme provably provides improved Byzantine fault-tolerance compared to other aggregation schemes in federated model. However, computing geometric median is a challenging problem, as there does not exist a closed-form formula~\cite{bajaj1988algebraic}. Moreover, existing numerical algorithms for computing geometric median are only approximate, and computationally quite complex~\cite{cohen2016geometric}. Other schemes, such as the verifiable coding in~\cite{so2020byzantine} and manual verification of information sent by agents~\cite{cao2020fltrust}, are not directly applicable to the commonly used federated framework where inter-agent communication is absent or there are a large number of agents, and data privacy is a major concern.

Besides federated local SGD framework, the proposed CE filter can also guarantee exact fault-tolerance in the distributed SGD framework where agents share their gradients instead of estimates (references omitted to preserve authors' anonymity). Also, for the distributed SGD framework, recent works have shown that {\em momentum} helps improve the fault-tolerance of a Byzantine-robust aggregation scheme~\cite{mhamdi2021distributed, sai2020learning}. However, adaptation of their results to federated local SGD method is non-trivial and remains to be investigated.

%!TEX root = main.tex

\section{Local SGD under Byzantine Model}\label{sec:alg}
We now present the proposed algorithm for solving \eqref{prob:obj} in the presence of at most $f$ Byzantine faulty agents. We note that the Byzantine agents can observe the values of other agents and send arbitrarily values to the coordinator.  To handle this scenario, the main idea of our approach is a Byzantine robust aggregation rule (or filter),  named comparative elimination (CE) filter, which is implemented at the coordinator. This filter together with the local SGD formulates our proposed method, formally presented in Algorithm \ref{alg:LocalSGD_CEfilter} for solving \eqref{prob:hon_obj}.    

In Algorithm \ref{alg:LocalSGD_CEfilter}, each agent $i$ maintains a local variable $x^{i}$, and the coordinator maintains $\xbar$, the average of these $x^{i}$. At any iteration $k\geq 0$, agent $i$ receives $\xbar_{k}$ from the coordinator and initializes its iterate $x^{i}_{k,0} = \xbar_{k}$. Here $x^{i}_{k,t}$ denotes the iterate at iteration $k$ and local time $t\in[0,\ldots,\Tcal-1]$ at agent $i$. Agent $i$ then runs a number $\Tcal$ of  local {SGD} steps using time-varying step sizes $\alpha_{k}$ and its local direction $G^{i}(x_{k,t}^{i})$, which can be either the {\em actual} gradient $\nabla q^{i}(x_{k,t}^{i})$ or a {\em stochastic} estimate $\nabla Q^{i}(\cdot,X^{i})$ of its gradient based on the data $\{X^{i}_{k,t}\}$ sampled i.i.d from $\pi^{i}$. After $\Tcal$ local {SGD} steps, the agents then send their new local updates $x^{i}_{k,\Tcal}$ to the coordinator. However, Byzantine agents may send arbitrary values to disrupt the learning process. The coordinator implements the CE filter (in steps $2(a)$ and $2(b)$ of Algorithm \ref{alg:LocalSGD_CEfilter}) to dilute the impact of ``bad" values sent by the Byzantine agents. The main of this filter is to discard $f$-values (or estimates) that are $f$-farthest from the current global estimate $\xbar_{k}$. Finally, the coordinator averages the $N-f$ remaining estimates, as shown in \eqref{alg:xbar}, to compute the new global estimate. Note that without the CE filter (i.e., without steps $2(a)$ and $2(b)$, and $F_{k} = [1,N]$), Algorithm \ref{alg:LocalSGD_CEfilter} reduces to the traditional local GD method.       

\begin{algorithm}[t]
\textbf{Initialization:} 
The coordinator initializes $\xbar_{0} \in\Rset^{d}$. Agent $i$ initializes step sizes $\{\alpha_{k}\}$ and a positive integer $T$.\\ 
\textbf{Iterations}: For $k=0,1,2,...$\vspace{-0.1cm}
\begin{enumerate}[leftmargin = 5mm]
    \item Agent $i$ \vspace{-0.1cm}
\begin{enumerate}[leftmargin = 5mm]
    \item Receive $\xbar_{k}$ sent by the server and set $x^{i}_{k,0} = \xbar_{k}$
    \item For $t = 0,1,\ldots,\Tcal-1$, implement \vspace{-0.1cm}
    \begin{align}
        x^{i}_{k,t+1} 
        &= x^{i}_{k,t} - \alpha_{k}  G^{i}(x^{i}_{k,t}).\label{alg:x_i}\vspace{-0.1cm}
    \end{align}
\end{enumerate}    
\item The coordinator receives $x^{i}_{k,T}$ from each agent $i$ and implement the CE filter as follows.\vspace{-0.1cm} 
\begin{enumerate}[leftmargin = 5mm]
\item Compute the distances of $x_{k,t}^{i}$ with its current value $\xbar_{k}$, and sort them in an increasing order
\begin{align}
\|\xbar_{k} - x^{i_{1}}_{k,\Tcal}\| \leq \ldots\leq \|\xbar_{k} - x^{i_{N}}_{k,\Tcal}\|.\label{alg:CEfilter} 
\end{align}
\item Discard the $f$-largest distances, i.e., it drops $x^{i_{N-f+1}}_{k,\Tcal}, \ldots,x^{i_{N}}_{k,\Tcal}$. Let ${\Fcal_{k} = \{i_{1},\ldots,i_{N-f}\}}$. 
\item Update its iterate as \vspace{-0.1cm}
  \begin{align}
      \xbar_{k+1} = \frac{1}{|\Fcal_{k}|}\sum_{i\in\Fcal_{k}}x^{i}_{k,\Tcal}. \label{alg:xbar}\vspace{-0.4cm}
  \end{align}
\end{enumerate}
 \end{enumerate}
\caption{Local SGD with CE Filter}
\label{alg:LocalSGD_CEfilter}
\end{algorithm}

%!TEX root = main.tex

\section{Main Results}\label{sec:results}
In this section, we present the main results of this paper, where we characterize the convergence of Algorithm \ref{alg:LocalSGD_CEfilter} for solving problem \eqref{prob:hon_obj}. We consider two cases, namely, the deterministic settings (when $G^{i}(\cdot) = \nabla q^{i}(\cdot))$ and the stochastic settings (when $G^{i}(\cdot) = \nabla Q^{i}(\cdot,X^{i})$).\footnote{Proofs of all the theorems presented in this section are deferred to the appendix attached after the list of references.} In both cases, our theoretical results are derived when the non-faulty agents' cost functions are smooth. Moreover, we also assume the average non-faulty cost function, denoted by $q^{\Hcal}(x)$, to be strongly convex. Specifically,
\begin{align}
    q^{\Hcal}(x) = \frac{1}{|\Hcal|} \sum_{i \in \Hcal}q^i(x).\label{eqn:def_cost_h}
\end{align}
These assumptions are formally stated as follows. 
% \begin{assumption}[Existence]
% \label{asp:finite}
% We assume non-trivial existence of a solution~\eqref{eqn:hon_obj}. Specifically, there exists a point
% \[x^*_{\H} \in \arg \min_{x \in \R^d} \sum_{i \in \H} Q_i(x) \text{ such that } \norm{x^*_\H} < \infty.\]
% \end{assumption}
% ~
\begin{assump}[Lipschitz smoothness]
\label{asp:lipschitz}
The non-faulty agents' functions have Lispchitz continuous gradients, i.e., there exists a positive constant $L < \infty$ such that, $\forall i \in \Hcal$,
\begin{align*}
    \|\nabla q^i(x) - \nabla q^i(y)\| \leq L \|x - y\|, \quad \forall x, \, y \in \Rset^d.
\end{align*}
\end{assump}
\begin{assump}[Strong convexity]
\label{asp:str_cvxty}
$q^{\Hcal}$ is strongly convex, i.e., there exists a positive constant $\mu < \infty$ such that
\begin{align*}
    (x - y)^T \left(\nabla q^{\Hcal}(x) - \nabla q^{\Hcal}(y) \right) \geq \mu \|x - y\|^2, \; \forall x, \, y \in \Rset^d
\end{align*}
where $(\cdot)^T$ denotes the transpose.
\end{assump}
To this end, we assume that these assumptions and the $2f$-redundancy property always hold true. In addition, without loss of generality we consider $|\Hcal| = N - f$. Finally, note that Assumptions~\ref{asp:lipschitz} and~\ref{asp:str_cvxty} hold true simultaneously only if $\mu \leq L$. 

\begin{remark}
Assumption \ref{asp:str_cvxty} implies that there exists a unique solution $x_{\Hcal}^{\star}$ of problem \eqref{prob:hon_obj}. However, this assumption does not imply that each local function $q^{i}$ is strongly convex. Indeed, each $q^{i}$ can have more than one minimizer. Under the $2f$-redundancy property one can show that 
\begin{align}
    x_{\Hcal}^{\star} \in \bigcap_{i \in \Hcal} \underset{x \in \Rset^d}{\arg\min} \, q^i(x). \label{eqn:int_min_sets}
\end{align}
% Moreover, when both Assumption~\ref{asp:lipschitz} and~\ref{asp:str_cvxty} hold true, along with the $2f$-redundancy property, then 
% \begin{align}
%     \lambda \leq \mu. \label{eqn:lambda_mu}
% \end{align}
Thus, one can view that Algorithm \ref{alg:LocalSGD_CEfilter} tries to search one point in the intersection of the minimizer sets of the local functions $q^{i}$. However, we do not assume we can compute these sets since this task is intractable in general. Finally, our analysis given later will rely on \eqref{eqn:int_min_sets}, whose proof can be found in \cite[Appendix B]{gupta2020fault_podc}. 
\end{remark}
\subsection{Deterministic Settings}\label{sec:deterministic}
In this section, we consider the deterministic setting of Algorithm \ref{alg:LocalSGD_CEfilter}, i.e., $G^{i}(\cdot) = \nabla q^{i}(\cdot)$. For convenience, we first study the convergence of Algorithm \ref{alg:LocalSGD_CEfilter} when $\Tcal=1$ in Section \ref{subsec:deterministic:T=1} and generalize to the case $\Tcal>1$ in Section \ref{subsec:deterministic:T>1}.

% Due to space limit, we only present the proof of the case $T=1$, which also help us to better present the main idea of our analysis. The analysis of other cases will be provided in the longer version of this paper.   

%!TEX root = main.tex

\subsubsection{The case of $\Tcal=1$}\label{subsec:deterministic:T=1}
When $\Tcal=1$, Algorithm \ref{alg:LocalSGD_CEfilter} is equivalent to the popular distributed (stochastic) gradient method. Indeed, by \eqref{alg:x_i} we have for any $i\in\Hcal$
\begin{align}
x^{i}_{k,1} = x^{i}_{k,0} - \alpha_{k}\nabla q^{i} (x^{i}_{k,0}) = \xbar_{k} - \alpha_{k}\nabla q^{i} (\xbar_{k}).\label{subsec:deterministic:T=1:xi}     
\end{align}
We denote by $\Bcal$ the set of Byzantine agents, i.e., $N = |\Bcal| + |\Hcal|$ and $|\Bcal| \leq f$. Without loss of generality we assume that $|\Bcal| = f$. Similarly, let $\Bcal_{k}$ be the set of Byzantine agents in $\Fcal_{k}$ and $\Hcal_{k}$ be the set of nonfaulty agents in $\Fcal_{k}$. Then we have $|\Bcal_{k}| = |\Fcal_{k}\setminus\Hcal_{k}| \leq f$, for any $k\geq 0$.

\begin{thm}\label{thm:deterministic:T=1}
Let $\{\xbar_{k}\}$ be generated by Algorithm \ref{alg:LocalSGD_CEfilter} with $\Tcal=1$. We assume that the following condition holds 
\begin{align}
\frac{f}{N-f} \leq \frac{\mu}{3L}\cdot\label{thm:deterministic:T=1:f_cond}
\end{align}
Let $\alpha_{k}$ be chosen as
\begin{align}
\alpha_{k} = \alpha \leq \frac{\mu}{4L^2}\cdot \label{thm:deterministic:T=1:stepsize}
\end{align}
Then we have 
\begin{align}
\|\xbar^{k} - \xbar_{\Hcal}^{\star}\|^2 \leq \left(1-\frac{\mu\alpha}{6} \right)^{k}\left\|\xbar_{0} - x_{\Hcal}^{\star}\right\|^2.    \label{thm:deterministic:T=1:ineq}
\end{align}
\end{thm}

\begin{remark}
In Theorem \ref{thm:deterministic:T=1} we show that under the $2f$ redundancy, Algorithm \ref{alg:LocalSGD_CEfilter} returns an exact solution $x_{\Hcal}^{\star}$ of problem \eqref{prob:hon_obj} even under of at most $f$ Byzantine agents. Moreover, the convergence is linear, which is the same as what we expect in the non-faulty case (no Byzantine agents).  
\end{remark}

\subsubsection{The case of $\Tcal>1$}\label{subsec:deterministic:T>1}
We now generalize Theorem \ref{thm:deterministic:T=1} to the case $\Tcal>1$, i.e., each agent implements more than $1$ local GD steps. This is indeed a common practice in federated optimization. When $\Tcal>1$, by \eqref{alg:x_i} we have  $\forall i\in\Hcal$ and $t\in[0,\Tcal)$
\begin{align}
x^{i}_{k,t+1} = \xbar_{k} -\alpha_{k}\sum_{\ell = 0}^{t}\nabla q^{i}(x^{i}_{k,\ell})\label{subsec:deterministic:T>1:xi},    
\end{align}
\begin{thm}\label{thm:deterministic:T>1}
Assume that \eqref{thm:deterministic:T=1:f_cond} hold and let 
$\alpha_{k}$ satisfy
\begin{align}
\alpha_{k} = \alpha \leq \frac{\mu}{16\Tcal L^2}\cdot \label{thm:deterministic:T>1:stepsize}
\end{align}
Then we have 
\begin{align}
\|\xbar^{k} - \xbar_{\Hcal}^{\star}\|^2 \leq \Big(1 - \frac{\mu \Tcal \alpha}{6}\Big)^{k}\left\|\xbar_{0} - x_{\Hcal}^{\star}\right\|^2.    \label{thm:deterministic:T>1:ineq}
\end{align}
\end{thm}

\begin{figure*}[t]
\centering
\begin{tabular}{cc}
\includegraphics[width=0.45\linewidth]{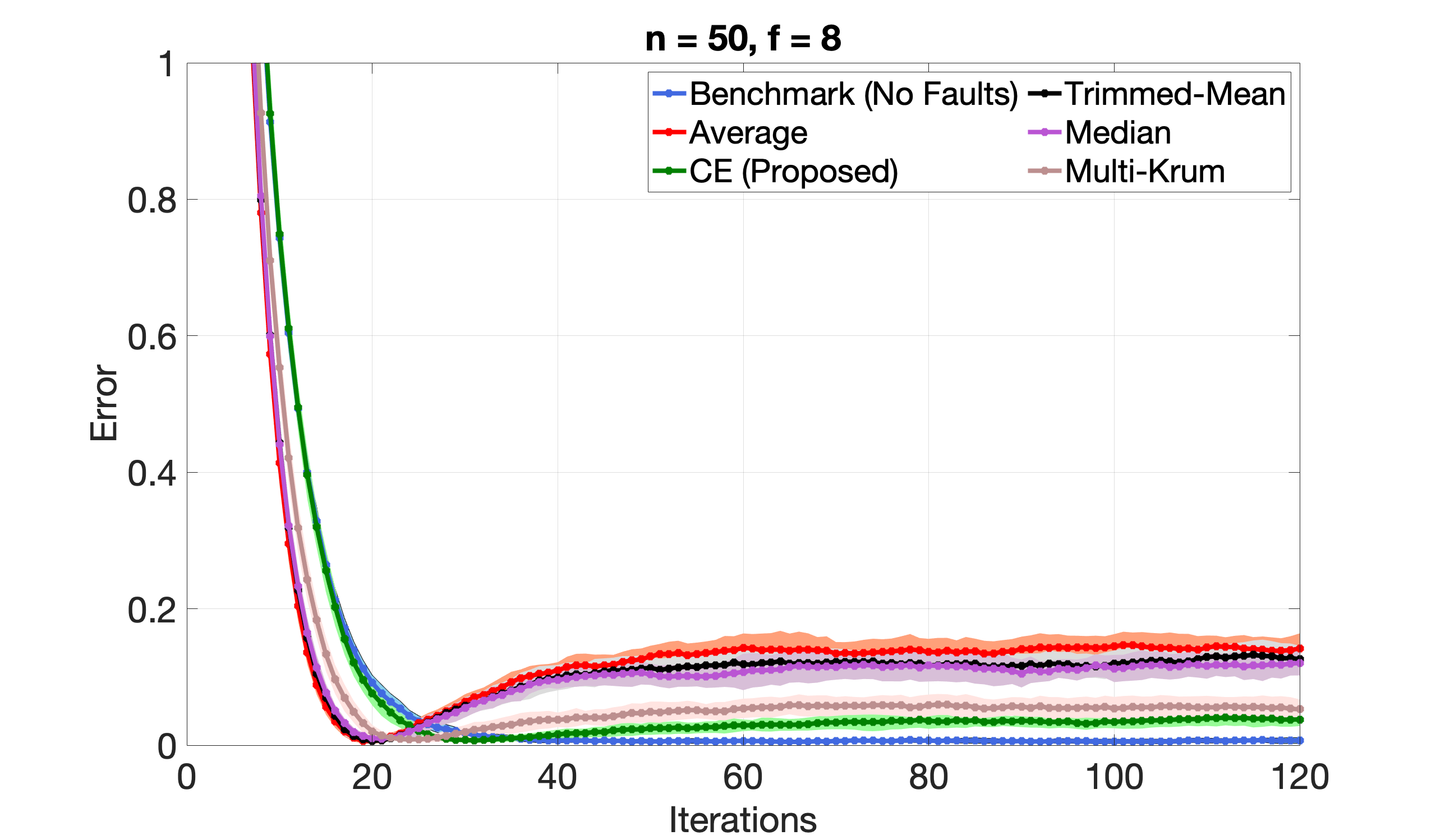}
&
\includegraphics[width=0.45\linewidth]{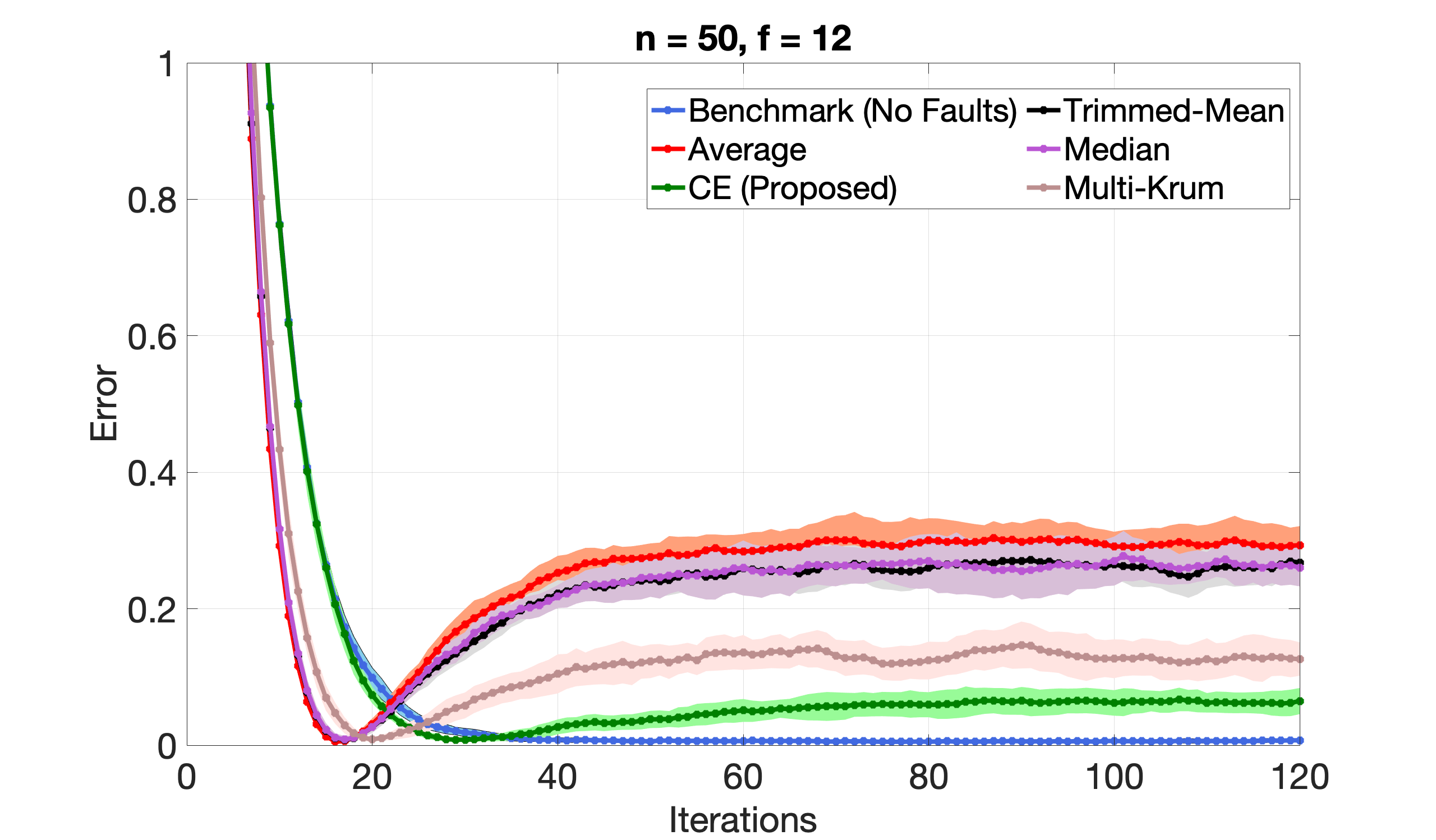}\\

\includegraphics[width=0.45\linewidth]{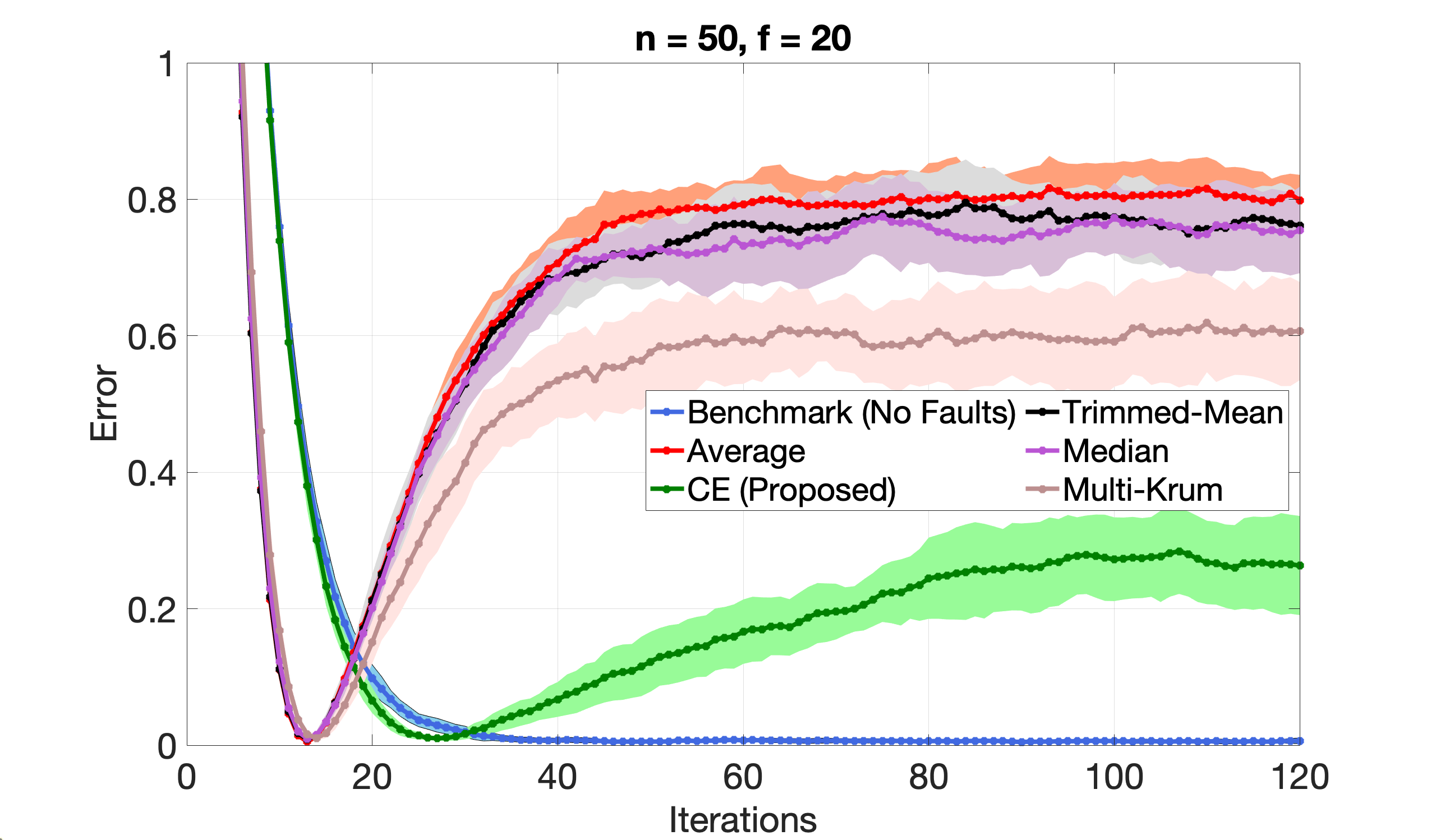}
&
\includegraphics[width=0.45\linewidth]{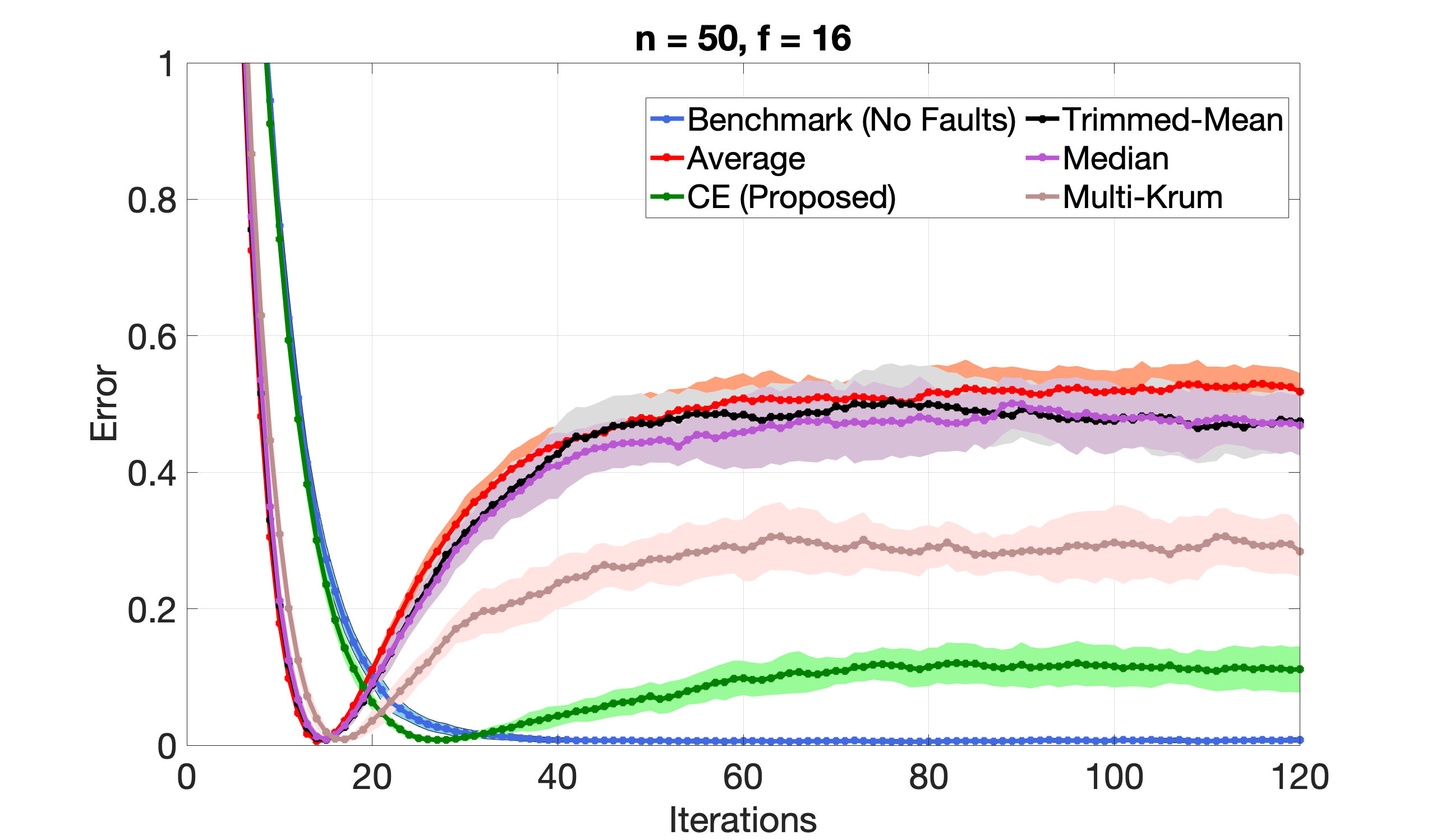}
\end{tabular}
\caption{The plots show the {\em error} $\|\xbar_k - x^*\|^2$ in {\em iteration} $k = 1, \ldots, \, 120$ of local GD (cf.~Algorithm~\ref{alg:LocalSGD_CEfilter}) with four different aggregation schemes; {\em averaging}, {\em CE}, {\em multi-KRUM}, {\em CWTM}, and {\em coordinate-wise median}. The benchmark corresponds to the fault-free execution of local GD. Solid lines show the mean performances of the schemes, and the shadows show the variance of their performances, observed over $100$ runs. In the clockwise order, $f = 8, \, 12, \, 16$ and $20$. We observe that, expectedly, all schemes obtain improved accuracy in presence of fewer Byzantine agents. We also observe that the performance of CE filter is consistently better than other schemes.}
\label{fig:diff_f}
\end{figure*}

\begin{figure*}[htb!]
\centering
\begin{tabular}{ccc}

\includegraphics[width=0.32\linewidth]{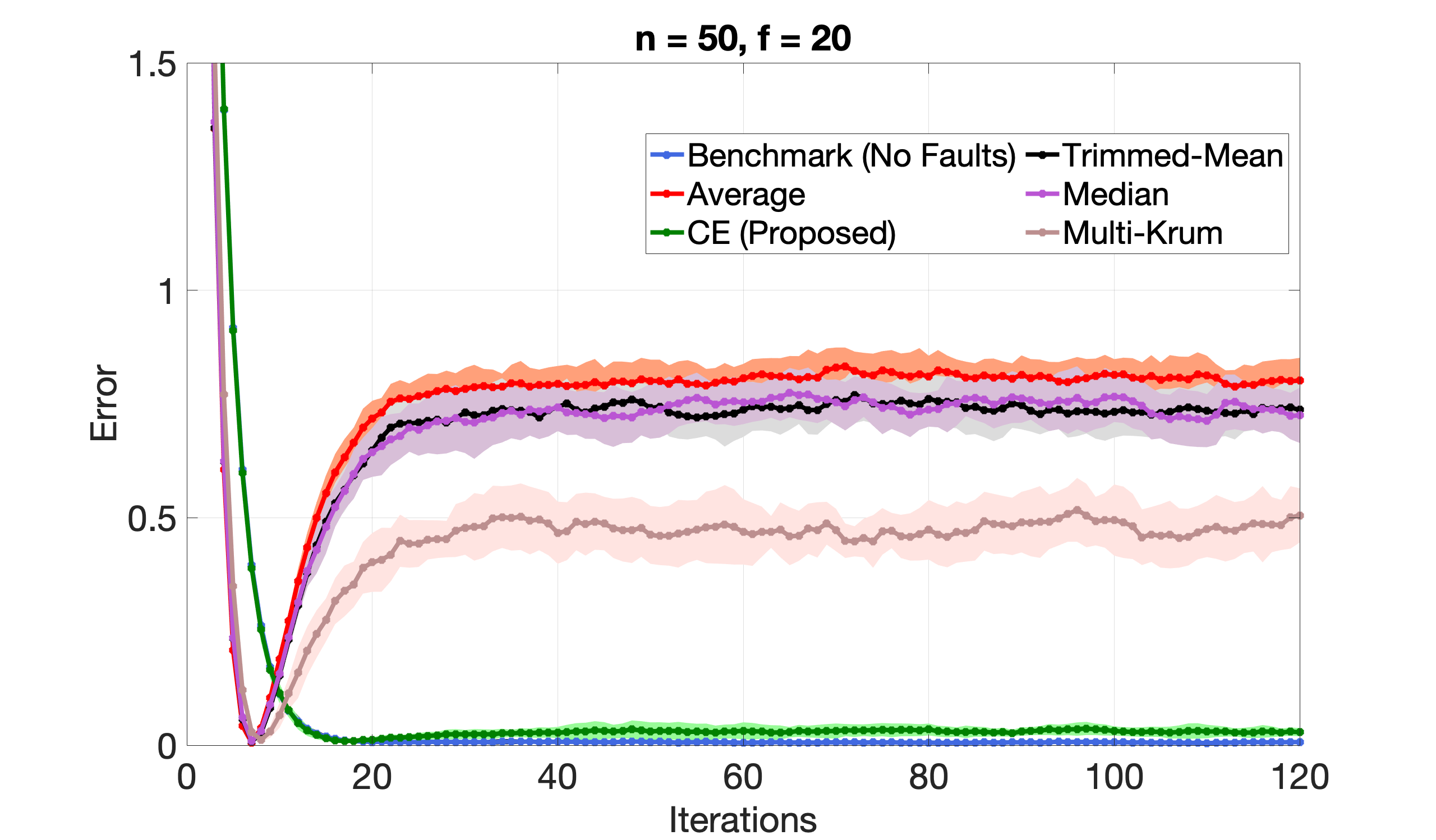}

\includegraphics[width=0.32\linewidth]{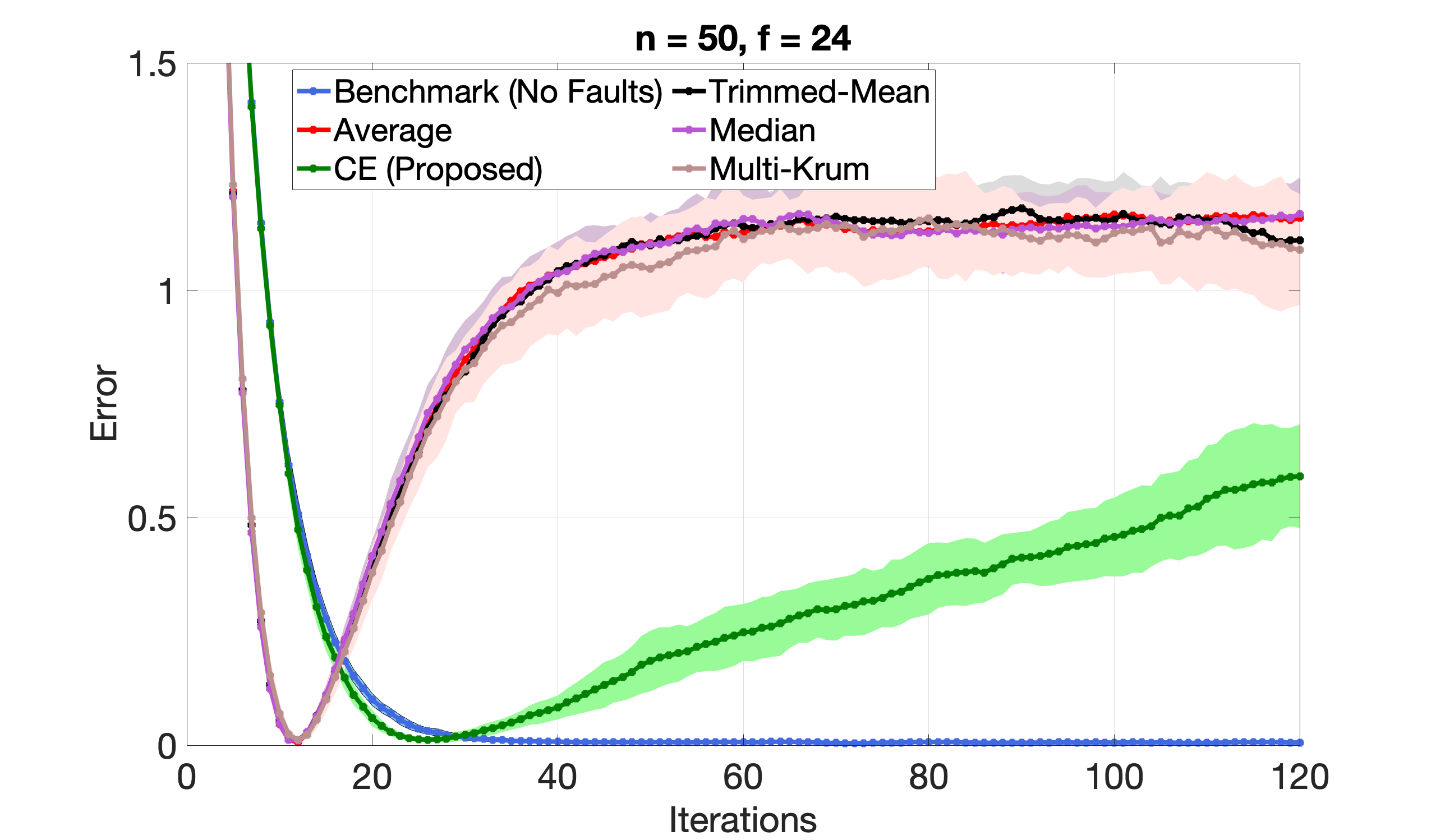}

\includegraphics[width=0.32\linewidth]{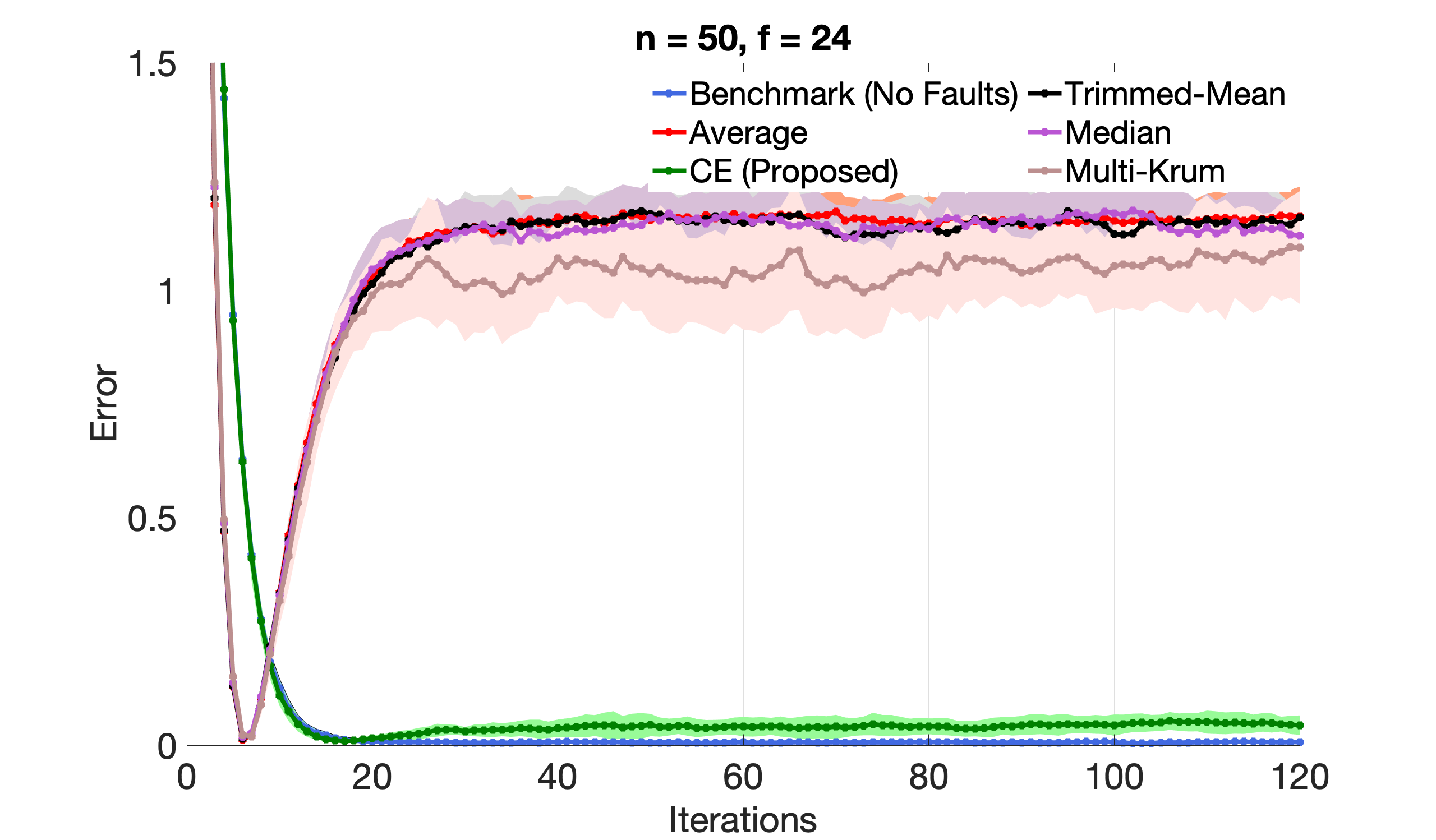}
\end{tabular}
\caption{The leftmost plot is for $f = 20$ and $T = 2$, the case with $T = 1$ is shown in Figure~\ref{fig:diff_f}. The middle and the rightmost plots are for $(f = 24, \, T = 1)$ and $(f = 24, \, T = 2)$, respectively. We observe that the accuracy of the local GD method with CE filter improves considerably as the number of local GD steps $T$ is increased from $1$ to $2$. The same cannot be observed for other schemes.}
\label{fig:diff_t}
\end{figure*}

%!TEX root = main.tex

\subsection{Stochastic Settings }\label{subsec:stochastic}
We next consider the setting where each agent only has access to the samples of its gradient, i.e., $G^{i}(\cdot) =  \nabla Q^{i}(\cdot,X^{i})$, where $X^{i}$ is a sequence of random variables sampled i.i.d from $\pi^{i}$. In the sequel, we denote by 
\[\Pcal_{k,t} = \cup_{i\in\Hcal}\{\xbar^{0},\ldots,\xbar_{k},x^{i}_{k,1},\ldots,x^{i}_{k,t}\}\] 
the filtration  containing all the history generated by Algorithm \ref{alg:LocalSGD_CEfilter} up to time $k+t$. To study the convergence of Algorithm \ref{alg:LocalSGD_CEfilter} we consider the following assumptions, which is often assumed in the literature of stochastic federated optimization \cite{Kairouz_survery2020}.

\begin{assump}\label{assump:noise}
The random variables $X_{k}^{i}$, for all $i$ and $k$, are i.i.d., and there exists a positive constant $\sigma$ such that
\begin{align*}
&\Eset[\nabla Q^{i}(x,X_{k,t}^{i})\,|\,\Pcal_{k,t}] = \nabla q^{i}(x),\quad \forall x\in\Rset^{d},\\ 
&\Eset[\|\nabla Q^{i}(x,X_{k,t}^{i}) - \nabla q^{i}(x)\|^2\,|\,\Pcal_{k,t}] \leq \sigma^2,\quad \forall x\in\Rset^{d}.
\end{align*}
\end{assump}
% \begin{assump}\label{assump:noise:Lipschitz}
% There exists a constant $B>0$ such that 
% \begin{align*}
% &\|\nabla Q^{i}(x,X^{i})\|\leq B (\|x\|+1)\quad\text{a.s.}
% \end{align*}
% \end{assump}
% Note that since the statistical sample space $\Xcal_{i}$ is compact, Assumption \ref{assump:noise:Lipschitz} is equivalent to assume that $\nabla Q^{i}(\cdot,X^{i})$ is Lipschitz continuous given any $X^{i}\in\Xcal^{i}$. This assumption is weaker than the uniform boundedness condition on $\nabla Q^{i}(\cdot,\cdot)$. 

Recall that $|\Bcal_{k}| + |\Hcal_{k}| = |\Fcal_{k}| = |\Hcal|$, for any $k\geq 0$. Finally, for convenience we denote by  
\begin{align*}
\nabla Q^{i}(x;X) = \frac{1}{|\Hcal|}\sum_{i\in\Hcal}\nabla Q^{i}(x;X^{i}),    
\end{align*}
where $X = (X^{1},\ldots,X^{|\Hcal|})^{T}$. 

\subsubsection{The case of $\Tcal=1$}
\begin{thm}
\label{thm:stochastic:T=1}
Suppose that Assumptions \ref{assump:noise} and condition \eqref{thm:deterministic:T=1:f_cond} hold. Moreover, let 
$\alpha_{k}$ be chosen as
\begin{align}
\alpha_{k} = \alpha \leq \frac{\mu}{12L^2}\cdot \label{thm:stochastic:T=1:stepsize}
\end{align}
Then we have 
\begin{align}
&\Eset[\|\xbar_{k} - x_{\Hcal}^{\star}\|^2]\notag\\ 
&\hspace{-0.2cm}\leq \left(1-\frac{\mu}{6}\alpha\right)^{k}\Eset[\|\xbar_{0} - x_{\Hcal}^{\star}\|^2] + \frac{14\sigma^2\alpha}{\mu} + \frac{2\sigma^2f}{\mu L|\Hcal|}\cdot   \label{thm:stochastic:T=1:ineq}
\end{align}
\end{thm}

\begin{remark}
Note that in Theorem \ref{thm:stochastic:T=1} due to the constant step size, the mean square error converges linearly only to a ball centered at the origin. The size of this ball depends on two terms: 1) one depends on the step size $\alpha$ which often seen in the convergence of local GD with non-faulty agents and 2) the other depends on the level of the gradient noise (or $\sigma$). The latter is due to the impact of the Byzantine agents and the stochastic gradient samples. Indeed, our comparative filter is designed to remove the potential bad values sent by the Byzantine agents, but not the variance of their stochastic samples. A potential solution for this issue is to let each agent sample a mini-batch of size $m$. In which case, $\sigma^2$ in \eqref{thm:stochastic:T=1:ineq} is replaced by $\sigma^2/m$. Thus, one can choose $m$ large enough so that the mean square error can get arbitrarily close to zero. Finally, when $\alpha_{k} \sim 1/k$, we can show that the convergence rate is sublinear $\Ocal(1/k)$.            
\end{remark}

\subsubsection{The case of $\Tcal>1$}
We next generalize Theorem~\ref{thm:stochastic:T=1} to the case when each agent implements more than $1$ local SGD steps, i.e.,$\Tcal>1$. By~\eqref{alg:x_i}, for all $i\in\Hcal$ and $t\in[0,\Tcal)$,
\begin{align}
x^{i}_{k,t+1} = \xbar_{k} -\alpha_{k}\sum_{\ell = 0}^{t}\nabla Q^{i}(x^{i}_{k,\ell};X_{k,\ell}^{i})\label{subsec:stochastic:T>1:xi}.
\end{align}
\begin{thm}
\label{thm:stochastic:T>1}
Suppose that Assumption \ref{assump:noise} and condition \eqref{thm:deterministic:T=1:f_cond} hold true. Moreover, let 
$\alpha_{k}$ be chosen as
\begin{align}
\alpha_{k} = \alpha \leq \frac{\mu}{144\Tcal L^2}\cdot \label{thm:stochastic:T>1:stepsize}
\end{align}
Then we have 
\begin{align}
\Eset[\|\xbar_{k} - x_{\Hcal}^{\star}\|^2]&\leq \Big(1-\frac{\mu \Tcal\alpha}{18}\Big)^{k}\Eset[\big\|\xbar_{0} - x_{\Hcal}^{\star}\big\|^2] \notag\\ 
&\quad + \frac{162\sigma^2 \Tcal}{\mu}\alpha + \frac{54\sigma^2f}{\mu L|\Hcal|}\cdot   \label{thm:stochastic:T=1:ineq}
\end{align}
\end{thm}

\section{Experiments}
\label{sec:sim}

To evaluate the efficacy of our proposed scheme, we simulate the problem of {\em robust mean estimation} in the federated framework. This problem serves as a test-case to empirically compare our scheme with others of similar computational costs, namely multi-KRUM~\cite{blanchard2017machine}, CWTM~\cite{su2018finite, yin2018byzantine}, and coordinate-wise median~\cite{xie2018generalized, yin2018byzantine}. For our experiments, we consider $N = 50$ agents and varying number of Byzantine faulty agents. Each non-faulty agent $i$ has $100$ noisy observations of a $10$-dimensional vector $x^*$ with all elements of unit value. In particular, the sample set $\mathcal{X}^i$ comprises $100$ uniformly distributed samples with each sample $X^i = x^* + Z^i$ where $Z^i \sim \mathcal{N}(0, I_d)$, and $Q^i(x;X^i) = (1/2) \|x - X^i\|^2$. In this case, $x^*$ is the unique solution to problem~\eqref{prob:hon_obj} for any set of honest agents $\mathcal{H}$. In our experimental settings, a Byzantine faulty agent $j$ behaves just like an honest agent with $100$ uniformly distributed samples, however each of its sample $X^j = 2 \times x^* + Z^j$ where $Z^j \sim \mathcal{N}(0, I_d)$. That is, honest agents send information corresponding to Gaussian noisy observations of $x^*$ and Byzantine 
agents send information corresponding to Gaussian noisy observations (with identical variance) of $2 \times x^*$. 

We simulate the {\em stochastic setting} of local GD (cf.~Algorithm~\ref{alg:LocalSGD_CEfilter}) with different number of faulty agents $f \in \{8, 12, 16, 20, 24\}$, different values of $\Tcal \in \{1, 2\}$, and different aggregation schemes in Step 2: {\em CE}, {\em mutli-KRUM}, {\em CWTM}, {\em coordinate-wise median} and {\em simple averaging}. The step-size $\alpha_k = 0.1$ for all $k$. Each setting is run $100$ times, and the observed errors $\|\xbar_k - x^*\|^2$ for $k = 1, \ldots, \, 120$ are shown in Figures~\ref{fig:diff_f} and~\ref{fig:diff_t}.
% ; solid lines and their shadows (with matching colors) represent the mean and the variance of the respective measurements.

{\bf Conclusion:} As suggested from our theoretical results, the final error upon using CE aggregation scheme decreases with the fraction of Byzantine faulty agents. We observe that CE aggregation scheme performs consistently better than multi-KRUM, CWTM and median. Moreover, we also observe that increasing the number of local gradient-descent steps, i.e., $\Tcal$, improves the fault-tolerance of CE aggregation scheme. However, the same cannot be said for other schemes. 

\section*{Summary}
In this paper, we have considered the problem of Byzantine fault-tolerance in the federated local stochastic gradient-descent method. We have proposed a new aggregation scheme, named comparative elimination (CE), and studied its fault-tolerance properties in both deterministic and stochastic settings. In the deterministic setting, we have shown the CE filter guarantees {\em exact fault-tolerance} against a bounded fraction of Byzantine agents $f/N$, provided the non-faulty agents' costs satisfy the necessary condition of $2f$-redundancy. In the stochastic setting, we have shown that CE filter obtains {\em approximate} fault-tolerance where the approximation error is proportional to the variance of the agents' stochastic gradients and the fraction of Byzantine agents.

\bibliographystyle{IEEEtran}
\bibliography{refs}

\begin{thebibliography}{10}
\providecommand{\url}[1]{#1}
\csname url@rmstyle\endcsname
\providecommand{\newblock}{\relax}
\providecommand{\bibinfo}[2]{#2}
\providecommand\BIBentrySTDinterwordspacing{\spaceskip=0pt\relax}
\providecommand\BIBentryALTinterwordstretchfactor{4}
\providecommand\BIBentryALTinterwordspacing{\spaceskip=\fontdimen2\font plus
\BIBentryALTinterwordstretchfactor\fontdimen3\font minus
  \fontdimen4\font\relax}
\providecommand\BIBforeignlanguage[2]{{%
\expandafter\ifx\csname l@#1\endcsname\relax
\typeout{** WARNING: IEEEtran.bst: No hyphenation pattern has been}%
\typeout{** loaded for the language `#1'. Using the pattern for}%
\typeout{** the default language instead.}%
\else
\language=\csname l@#1\endcsname
\fi
#2}}

\bibitem{Kairouz_survery2020}
P.~Kairouz and H.~B. McMahan, ``Advances and open problems in federated
  learning,'' \emph{Foundations and Trends® in Machine Learning}, vol.~14,
  no.~1, 2021.

\bibitem{Li_survery2020}
T.~{Li}, A.~K. {Sahu}, A.~{Talwalkar}, and V.~{Smith}, ``Federated learning:
  Challenges, methods, and future directions,'' \emph{IEEE Signal Processing
  Magazine}, vol.~37, no.~3, pp. 50--60, 2020.

\bibitem{lamport1982byzantine}
L.~Lamport, R.~Shostak, and M.~Pease, ``The {Byzantine} generals problem,''
  \emph{ACM Transactions on Programming Languages and Systems (TOPLAS)},
  vol.~4, no.~3, pp. 382--401, 1982.

\bibitem{xie2019dba}
C.~Xie, K.~Huang, P.-Y. Chen, and B.~Li, ``Dba: Distributed backdoor attacks
  against federated learning,'' in \emph{International Conference on Learning
  Representations}, 2019.

\bibitem{su2016fault}
L.~Su and N.~H. Vaidya, ``Fault-tolerant multi-agent optimization: optimal
  iterative distributed algorithms,'' in \emph{Proceedings of the 2016 ACM
  symposium on principles of distributed computing}.\hskip 1em plus 0.5em minus
  0.4em\relax ACM, 2016, pp. 425--434.

\bibitem{gupta2020fault_podc}
N.~Gupta and N.~H. Vaidya, ``Fault-tolerance in distributed optimization: The
  case of redundancy,'' in \emph{The 39th Symposium on Principles of
  Distributed Computing}, 2020, pp. 365--374.

\bibitem{gupta2020resilience}
------, ``Resilience in collaborative optimization: Redundant and independent
  cost functions,'' \emph{arXiv preprint arXiv:2003.09675}, 2020.

\bibitem{chong2015observability}
M.~S. Chong, M.~Wakaiki, and J.~P. Hespanha, ``Observability of linear systems
  under adversarial attacks,'' in \emph{American Control Conference}.\hskip 1em
  plus 0.5em minus 0.4em\relax IEEE, 2015, pp. 2439--2444.

\bibitem{gupta2019byzantine}
N.~Gupta and N.~H. Vaidya, ``{Byzantine} fault tolerant distributed linear
  regression,'' \emph{arXiv preprint arXiv:1903.08752}, 2019.

\bibitem{mishra2016secure}
S.~Mishra, Y.~Shoukry, N.~Karamchandani, S.~N. Diggavi, and P.~Tabuada,
  ``Secure state estimation against sensor attacks in the presence of noise,''
  \emph{IEEE Transactions on Control of Network Systems}, vol.~4, no.~1, pp.
  49--59, 2016.

\bibitem{su2019finite}
L.~Su and S.~Shahrampour, ``Finite-time guarantees for byzantine-resilient
  distributed state estimation with noisy measurements,'' \emph{IEEE
  Transactions on Automatic Control}, vol.~65, no.~9, pp. 3758--3771, 2019.

\bibitem{alistarh2018byzantine}
D.~Alistarh, Z.~Allen-Zhu, and J.~Li, ``Byzantine stochastic gradient
  descent,'' in \emph{Proceedings of the 32nd International Conference on
  Neural Information Processing Systems}, 2018, pp. 4618--4628.

\bibitem{blanchard2017machine}
P.~Blanchard, R.~Guerraoui, \emph{et~al.}, ``Machine learning with adversaries:
  {Byzantine} tolerant gradient descent,'' in \emph{Advances in Neural
  Information Processing Systems}, 2017, pp. 119--129.

\bibitem{charikar2017learning}
M.~Charikar, J.~Steinhardt, and G.~Valiant, ``Learning from untrusted data,''
  in \emph{Proceedings of the 49th Annual ACM SIGACT Symposium on Theory of
  Computing}, 2017, pp. 47--60.

\bibitem{guerraoui2018hidden}
R.~Guerraoui, S.~Rouault, \emph{et~al.}, ``The hidden vulnerability of
  distributed learning in byzantium,'' in \emph{International Conference on
  Machine Learning}.\hskip 1em plus 0.5em minus 0.4em\relax PMLR, 2018, pp.
  3521--3530.

\bibitem{liu2021approximate}
S.~Liu, N.~Gupta, and N.~H. Vaidya, ``Approximate byzantine fault-tolerance in
  distributed optimization,'' \emph{arXiv preprint arXiv:2101.09337}, 2021.

\bibitem{kuwaranancharoen2020byzantine}
K.~Kuwaranancharoen, L.~Xin, and S.~Sundaram, ``Byzantine-resilient distributed
  optimization of multi-dimensional functions,'' in \emph{2020 American Control
  Conference (ACC)}.\hskip 1em plus 0.5em minus 0.4em\relax IEEE, 2020, pp.
  4399--4404.

\bibitem{xie2020fall}
\BIBentryALTinterwordspacing
C.~Xie, O.~Koyejo, and I.~Gupta, ``Fall of empires: Breaking byzantine-tolerant
  sgd by inner product manipulation,'' in \emph{Proceedings of The 35th
  Uncertainty in Artificial Intelligence Conference}, ser. Proceedings of
  Machine Learning Research, R.~P. Adams and V.~Gogate, Eds., vol. 115.\hskip
  1em plus 0.5em minus 0.4em\relax PMLR, 22--25 Jul 2020, pp. 261--270.
  [Online]. Available: \url{https://proceedings.mlr.press/v115/xie20a.html}
\BIBentrySTDinterwordspacing

\bibitem{yin2018byzantine}
D.~Yin, Y.~Chen, K.~Ramchandran, and P.~Bartlett, ``Byzantine-robust
  distributed learning: Towards optimal statistical rates,'' in
  \emph{International Conference on Machine Learning}, 2018, pp. 5636--5645.

\bibitem{yang2019byrdie}
Z.~Yang and W.~U. Bajwa, ``Byrdie: Byzantine-resilient distributed coordinate
  descent for decentralized learning,'' \emph{IEEE Transactions on Signal and
  Information Processing over Networks}, vol.~5, no.~4, pp. 611--627, 2019.

\bibitem{chen2017distributed}
Y.~Chen, L.~Su, and J.~Xu, ``Distributed statistical machine learning in
  adversarial settings: Byzantine gradient descent,'' \emph{Proceedings of the
  ACM on Measurement and Analysis of Computing Systems}, vol.~1, no.~2, pp.
  1--25, 2017.

\bibitem{sundaram2018distributed}
S.~Sundaram and B.~Gharesifard, ``Distributed optimization under adversarial
  nodes,'' \emph{IEEE Transactions on Automatic Control}, 2018.

\bibitem{xie2018phocas}
\BIBentryALTinterwordspacing
C.~Xie, O.~Koyejo, and I.~Gupta, ``Phocas: dimensional byzantine-resilient
  stochastic gradient descent,'' \emph{CoRR}, vol. abs/1805.09682, 2018.
  [Online]. Available: \url{http://arxiv.org/abs/1805.09682}
\BIBentrySTDinterwordspacing

\bibitem{li2019rsa}
L.~Li, W.~Xu, T.~Chen, G.~B. Giannakis, and Q.~Ling, ``Rsa: Byzantine-robust
  stochastic aggregation methods for distributed learning from heterogeneous
  datasets,'' in \emph{Proceedings of the AAAI Conference on Artificial
  Intelligence}, vol.~33, 2019, pp. 1544--1551.

\bibitem{sohn2020election}
J.-y. Sohn, D.-J. Han, B.~Choi, and J.~Moon, ``Election coding for distributed
  learning: Protecting signsgd against byzantine attacks,'' \emph{Advances in
  Neural Information Processing Systems}, vol.~33, 2020.

\bibitem{diakonikolas2019sever}
\BIBentryALTinterwordspacing
I.~Diakonikolas, G.~Kamath, D.~Kane, J.~Li, J.~Steinhardt, and A.~Stewart,
  ``Sever: A robust meta-algorithm for stochastic optimization,'' in
  \emph{Proceedings of the 36th International Conference on Machine Learning},
  ser. Proceedings of Machine Learning Research, K.~Chaudhuri and
  R.~Salakhutdinov, Eds., vol.~97.\hskip 1em plus 0.5em minus 0.4em\relax PMLR,
  09--15 Jun 2019, pp. 1596--1606. [Online]. Available:
  \url{http://proceedings.mlr.press/v97/diakonikolas19a.html}
\BIBentrySTDinterwordspacing

\bibitem{prasad2020robust}
A.~Prasad, A.~S. Suggala, S.~Balakrishnan, and P.~Ravikumar, ``Robust
  estimation via robust gradient estimation,'' \emph{Journal of the Royal
  Statistical Society: Series B (Statistical Methodology)}, vol.~82, no.~3, pp.
  601--627, 2020.

\bibitem{su2020byzantine}
L.~Su and N.~H. Vaidya, ``Byzantine-resilient multi-agent optimization,''
  \emph{IEEE Transactions on Automatic Control}, 2020.

\bibitem{yang2017byrdie}
Z.~Yang and W.~U. Bajwa, ``Byrdie: {Byzantine}-resilient distributed coordinate
  descent for decentralized learning,'' 2017.

\bibitem{fang2020local}
M.~Fang, X.~Cao, J.~Jia, and N.~Gong, ``Local model poisoning attacks to
  byzantine-robust federated learning,'' in \emph{29th $\{$USENIX$\}$ Security
  Symposium ($\{$USENIX$\}$ Security 20)}, 2020, pp. 1605--1622.

\bibitem{munoz2019byzantine}
L.~Mu{\~n}oz-Gonz{\'a}lez, K.~T. Co, and E.~C. Lupu, ``Byzantine-robust
  federated machine learning through adaptive model averaging,'' \emph{arXiv
  preprint arXiv:1909.05125}, 2019.

\bibitem{wu2020federated}
Z.~Wu, Q.~Ling, T.~Chen, and G.~B. Giannakis, ``Federated variance-reduced
  stochastic gradient descent with robustness to byzantine attacks,''
  \emph{IEEE Transactions on Signal Processing}, vol.~68, pp. 4583--4596, 2020.

\bibitem{bajaj1988algebraic}
C.~Bajaj, ``The algebraic degree of geometric optimization problems,''
  \emph{Discrete \& Computational Geometry}, vol.~3, no.~2, pp. 177--191, 1988.

\bibitem{cohen2016geometric}
M.~B. Cohen, Y.~T. Lee, G.~Miller, J.~Pachocki, and A.~Sidford, ``Geometric
  median in nearly linear time,'' in \emph{Proceedings of the forty-eighth
  annual ACM symposium on Theory of Computing}, 2016, pp. 9--21.

\bibitem{so2020byzantine}
J.~So, B.~G{\"u}ler, and A.~S. Avestimehr, ``Byzantine-resilient secure
  federated learning,'' \emph{IEEE Journal on Selected Areas in
  Communications}, 2020.

\bibitem{cao2020fltrust}
X.~Cao, M.~Fang, J.~Liu, and N.~Z. Gong, ``Fltrust: Byzantine-robust federated
  learning via trust bootstrapping,'' \emph{arXiv preprint arXiv:2012.13995},
  2020.

\bibitem{mhamdi2021distributed}
\BIBentryALTinterwordspacing
E.~M.~E. Mhamdi, R.~Guerraoui, and S.~Rouault, ``Distributed momentum for
  byzantine-resilient stochastic gradient descent,'' in \emph{International
  Conference on Learning Representations}, 2021. [Online]. Available:
  \url{https://openreview.net/forum?id=H8UHdhWG6A3}
\BIBentrySTDinterwordspacing

\bibitem{sai2020learning}
\BIBentryALTinterwordspacing
S.~P. Karimireddy, L.~He, and M.~Jaggi, ``Learning from history for byzantine
  robust optimization,'' \emph{CoRR}, vol. abs/2012.10333, 2020. [Online].
  Available: \url{https://arxiv.org/abs/2012.10333}
\BIBentrySTDinterwordspacing

\bibitem{su2018finite}
L.~Su and S.~Shahrampour, ``Finite-time guarantees for {Byzantine}-resilient
  distributed state estimation with noisy measurements,'' \emph{arXiv preprint
  arXiv:1810.10086}, 2018.

\bibitem{xie2018generalized}
C.~Xie, O.~Koyejo, and I.~Gupta, ``Generalized {Byzantine}-tolerant sgd,''
  \emph{arXiv preprint arXiv:1802.10116}, 2018.

\end{thebibliography}

\appendix

\section{Proofs}
\subsection{Proof of Theorem~\ref{thm:deterministic:T=1}}
\label{app:deterministic:T=1}

\begin{proof}
Using \eqref{alg:xbar} and \eqref{subsec:deterministic:T=1:xi} we have
\begin{align}
&\xbar_{k+1} = \frac{1}{|\Fcal_{k}|}\sum_{i\in\Fcal_{k}}x^{i}_{k,1}\notag\\ 
&=  \frac{1}{|\Hcal|}\Big[\sum_{i\in\Hcal}x^{i}_{k,1} + \sum_{i\in\Bcal_{k}}x^{i}_{k,1} - \sum_{i\in\Hcal\setminus\Hcal_{k}}x^{i}_{k,1}\Big]\notag\\  
&\stackrel{\eqref{subsec:deterministic:T=1:xi}}{=} \xbar_{k} - \frac{\alpha_{k}}{|\Hcal|}\sum_{i\in\Hcal}\nabla q^{i}(\xbar_{k})\notag\\ 
&\quad + \frac{1}{|\Hcal|}\Big[ \sum_{i\in\Bcal_{k}}x^{i}_{k,1} - \sum_{i\in\Hcal\setminus\Hcal_{k}}x^{i}_{k,1}\Big]\notag\\
&= \xbar_{k} - \alpha_{k}\nabla q^{\Hcal}(\xbar_{k}) \notag\\ 
&\quad + \frac{1}{|\Hcal|}\Big[ \sum_{i\in\Bcal_{k}}(x^{i}_{k,1} - \xbar_{k}) -\!\!\! \sum_{i\in\Hcal\setminus\Hcal_{k}}(x^{i}_{k,1}-\xbar_{k})\Big],\label{subsec:deterministic:T=1:xbar}
\end{align}
where the last equality is due to $|\Bcal_{k}| = |\Hcal\setminus\Hcal_{k}|$. Using the preceding relation we consider 
\begin{align}
&\|\xbar_{k+1} - x_{\Hcal}^{\star}\|^2\notag\\ 
&= \|\xbar_{k} - x_{\Hcal}^{\star} - \alpha_{k}\nabla q^{\Hcal}(\xbar_{k})\|^2\notag\\ 
&\quad + \Big\| \frac{1}{|\Hcal|} \sum_{i\in\Bcal_{k}}(x^{i}_{k,1} - \xbar_{k}) - \frac{1}{|\Hcal|}\sum_{i\in\Hcal\setminus\Hcal_{k}}(x^{i}_{k,1}-\xbar_{k}) \Big\|^2\notag\\
&\quad + \frac{2}{|\Hcal|}\sum_{i\in\Bcal_{k}}\left(\xbar_{k} - x_{\Hcal}^{\star} - \alpha_{k}\nabla q^{\Hcal}(\xbar_{k})\right)^T(x^{i}_{k,1} - \xbar_{k})\notag\\
&\quad - \frac{2}{|\Hcal|}\!\!\sum_{i\in\Hcal\setminus\Hcal_{k}}\!\!\!\!\!\left(\xbar_{k} - x_{\Hcal}^{\star} - \alpha_{k}\nabla q^{\Hcal}(\xbar_{k})\right)^T(x^{i}_{k,1} - \xbar_{k}).\label{subsec:deterministic:T=1:eq1}
\end{align}
We next analyze each term on the right-hand side of \eqref{subsec:deterministic:T=1:eq1}. First, using \eqref{subsec:deterministic:T=1:xi} we have for all $i\in\Hcal$
\begin{align*}
x^{i}_{k,1} - \xbar_{k} = -\alpha_{k}\nabla q^{i}(\xbar_{k}).    
\end{align*}
In addition, under $2$f-redundancy we have $x_{\Hcal}^{\star}\in \Xcal_i^{\star}$, where $\Xcal_{i}^{\star}$ is the set of minimizers of $q^{i}(x_{\Hcal}^{\star})$. This implies that $\nabla q^{i}(x_{\Hcal}^{\star}) = 0$. In addition, Assumption \ref{asp:lipschitz} implies that $\nabla q^{\Hcal}$ is also L-Lipschitz continuous. Recall that $|\Bcal_{k}| = |\Hcal\setminus\Hcal_{k}|$. Then, by Assumption \ref{asp:lipschitz} we consider the last term on the right-hand side of \eqref{subsec:deterministic:T=1:eq1}
\begin{align}
&-\frac{2}{|\Hcal|}  \sum_{i\in\Hcal\setminus\Hcal_{k}}(\xbar_{k} - x_{\Hcal}^{\star} - \alpha_{k}\nabla q^{\Hcal}(\xbar_{k}))^T(x^{i}_{k,1}-\xbar_{k})\notag\\ 
&= \frac{2\alpha_{k}}{|\Hcal|}  \sum_{i\in\Hcal\setminus\Hcal_{k}}(\xbar_{k} - x_{\Hcal}^{\star}-\alpha_{k}\nabla q^{\Hcal}(\xbar_{k}))^T\nabla q^{i}(\xbar_{k})\notag\\
&=  \frac{2\alpha_{k}}{|\Hcal|}  \sum_{i\in\Hcal\setminus\Hcal_{k}}\Big(\xbar_{k} - x_{\Hcal}^{\star}-\alpha_{k}(\nabla q^{\Hcal}(\xbar_{k})-\nabla q^{\Hcal}(x_{\Hcal}^{\star}))\Big)^T\notag\\
&\hspace{2.5cm} \times (\nabla q^{i}(\xbar_{k}) - \nabla q^{i}(x_{\Hcal}^{\star}))\notag\\
&\leq  \frac{2\alpha_{k}}{|\Hcal|}  \sum_{i\in\Hcal\setminus\Hcal_{k}}(\|\xbar_{k} - x_{\Hcal}^{\star}]\| + \alpha_{k}L\| \xbar_{k}- x_{\Hcal}^{\star}\|) L\|\xbar_{k} - x_{\Hcal}^{\star}\|\notag\\
&= \Big(\frac{2L|\Bcal_{k}|\alpha_{k}}{|\Hcal|}+\frac{2L^2|\Bcal_{k}|\alpha_{k}^2}{|\Hcal|}\Big)\|\xbar_{k}-x_{\Hcal}^{\star}\|^2.\label{subsec:deterministic:T=1:eq1a}
\end{align}
% Second, using Assumption \ref{asp:lipschitz} we consider
% \begin{align}
% &-\frac{2\alpha_{k}}{|\Hcal|}  \sum_{i\in\Hcal\setminus\Hcal_{k}}(-\nabla q^{\Hcal}(\xbar_{k}))^T(x^{i}_{k,1}-\xbar_{k})\notag\\
% &= -\frac{2\alpha_{k}^2}{|\Hcal|}  \sum_{i\in\Hcal\setminus\Hcal_{k}}\nabla q^{\Hcal}(\xbar_{k})^T\nabla q^{i}(\xbar_{k})\allowdisplaybreaks\notag\\  
% &= -\frac{2\alpha_{k}^2}{|\Hcal|}\!\!\! \sum_{i\in\Hcal\setminus\Hcal_{k}}\!\!\!\!(\nabla q^{\Hcal}(\xbar_{k})-\nabla q^{\Hcal}(x_{\Hcal}^{\star}))^T\notag\\
% &\hspace{2.5cm}\times\left(\nabla q^{i}(\xbar_{k})-\nabla q^{i}(x_{\Hcal}^{\star})\right)\notag\\
% &\leq \frac{2L^2|\Bcal_{k}|\alpha_{k}^2}{|\Hcal|}\|\xbar_{k}-x_{\Hcal}^{\star}\|^2.   \label{subsec:deterministic:T=1:eq1b}
% \end{align}
Second, using our CE filter (a.k.a \eqref{alg:CEfilter}) and \eqref{subsec:deterministic:T=1:xi}  there exists $j\in\Hcal$ such that for all $i\in\Bcal_{k}$ we have
\begin{align}
&\|x^{i}_{k,1}-\xbar_{k}\| \leq \|x^{j}_{k,1}-\xbar_{k}\| = \alpha_{k}\|\nabla q^{j}(\xbar_{k})\|\notag\\ 
& = \alpha_{k}\|\nabla q^{j}(\xbar_{k}) - \nabla q^{j}(x_{\Hcal}^{\star})\| \leq L\alpha_{k}\|\xbar_{k}-x_{\Hcal}^{\star} \|,\label{subsec:deterministic:T=1:eq1b1}
\end{align}
which implies that
\begin{align}
&\frac{2}{|\Hcal|}\sum_{i\in\Bcal_{k}}\left(\xbar_{k} - x_{\Hcal}^{\star} - \alpha_{k}\nabla q^{\Hcal}(\xbar_{k})\right)^T(x^{i}_{k,1} - \xbar_{k})\notag\\ 
&\leq \frac{2}{|\Hcal|}\sum_{i\in\Bcal_{k}}\left(\|\xbar_{k} - x_{\Hcal}^{\star}\| + \alpha_{k}\|\nabla q^{\Hcal}(\xbar_{k})\|\right)\|x^{i}_{k,1} - \xbar_{k}\|\notag\\
&= \frac{2}{|\Hcal|}\sum_{i\in\Bcal_{k}}\left(\|\xbar_{k} - x_{\Hcal}^{\star}\| + \alpha_{k}\|\nabla q^{\Hcal}(\xbar_{k})-\nabla q^{\Hcal}(x_{\Hcal}^{\star})\|\right)\notag\\
&\hspace{2.5cm}\times \|x^{i}_{k,1} - \xbar_{k}\|\notag\\
&\leq \Big(\frac{2L|\Bcal_{k}|\alpha_{k}}{|\Hcal|} + \frac{2L^2|\Bcal_{k}|\alpha_{k}^2}{|\Hcal|}\Big)\|\xbar_{k}-x_{\Hcal}^{\star}\|^2.  \label{subsec:deterministic:T=1:eq1b}
\end{align}
Next, using \eqref{subsec:deterministic:T=1:xi} and  \eqref{subsec:deterministic:T=1:eq1b1} we obtain 
\begin{align}
&\Big\| \frac{1}{|\Hcal|} \sum_{i\in\Bcal_{k}}(x^{i}_{k,1} - \xbar_{k}) - \frac{1}{|\Hcal|}\sum_{i\in\Hcal\setminus\Hcal_{k}}(x^{i}_{k,1}-\xbar_{k}) \Big\|^2\notag\\
&\leq \frac{2|\Bcal_{k}|}{|\Hcal|^2}\Big(  \sum_{i\in\Bcal_{k}}\|x^{i}_{k,1} - \xbar_{k}\|^2 + \sum_{i\in\Hcal\setminus\Hcal_{k}}\|x^{i}_{k,1}-\xbar_{k}\|^2\Big)\notag\\
&\leq \frac{4L^2|\Bcal_{k}|^2}{|\Hcal|^2}\alpha_{k}^2\|\xbar_{k}-x_{\Hcal}^{\star}\|^2.\label{subsec:deterministic:T=1:eq1d}
\end{align}
Finally, using Assumption \ref{asp:str_cvxty} we obtain
\begin{align}
&\left\|\xbar_{k} - x_{\Hcal}^{\star} - \alpha_{k}\nabla q^{\Hcal}(\xbar_{k})\right\|^2\notag\\ 
&= \|\xbar_{k} - x_{\Hcal}^{\star} \|^2 - 2\alpha_{k}\nabla q^{\Hcal}(\xbar_{k})^T(\xbar_{k} - x_{\Hcal}^{\star} )\notag\\ 
&\quad + \alpha_{k}^2\|\nabla q^{\Hcal}(\xbar_{k}) - \nabla q^{\Hcal}(x_{\Hcal}^{\star})\|^2\notag\\ 
&\leq (1-2\mu\alpha_{k} + L^2\alpha_{k}^2)\left\|\xbar_{k} - x_{\Hcal}^{\star}\right\|^2.   \label{subsec:deterministic:T=1:eq1e}
\end{align}
Substituting \eqref{subsec:deterministic:T=1:eq1a}--\eqref{subsec:deterministic:T=1:eq1e} into \eqref{subsec:deterministic:T=1:eq1} we obtain
\begin{align}
&\|\xbar_{k+1} - x_{\Hcal}^{\star}\|^2\notag\\ 
&\leq (1-2\mu\alpha_{k} + L^2\alpha_{k}^2)\left\|\xbar_{k} - x_{\Hcal}^{\star}\right\|^2\notag\\ 
&\quad +  \frac{4L^2|\Bcal_{k}|^2}{|\Hcal|^2}\alpha_{k}^2\|\xbar_{k}-x_{\Hcal}^{\star}\|^2\notag\\
&\quad + \Big(\frac{2L|\Bcal_{k}|\alpha_{k}}{|\Hcal|}+\frac{2L^2|\Bcal_{k}|\alpha_{k}^2}{|\Hcal|}\Big)\|\xbar_{k}-x_{\Hcal}^{\star}\|^2\notag\\
&\quad + \Big(\frac{2L|\Bcal_{k}|\alpha_{k}}{|\Hcal|} + \frac{2L^2|\Bcal_{k}|\alpha_{k}^2}{|\Hcal|}\Big)\|\xbar_{k}-x_{\Hcal}^{\star}\|^2\notag\\
&\leq (1-2\mu\alpha_{k} + L^2\alpha_{k}^2)\left\|\xbar_{k} - x_{\Hcal}^{\star}\right\|^2\notag\\ 
&\quad +  \frac{4L^2|\Bcal_{k}|}{|\Hcal|}\Big(1+\frac{|\Bcal_{k}|}{|\Hcal_{k}|}\Big)\alpha_{k}^2\|\xbar_{k}-x_{\Hcal}^{\star}\|^2\notag\\
&\quad + \frac{4L|\Bcal_{k}|\alpha_{k}}{|\Hcal|}\|\xbar_{k}-x_{\Hcal}^{\star}\|^2\notag\\
&\leq \Big(1-2\Big(\mu-\frac{2Lf}{|\Hcal|}\Big)\alpha_{k} \Big)\left\|\xbar_{k} - x_{\Hcal}^{\star}\right\|^2\notag\\ 
&\quad + \Big(1 +\frac{4f}{|\Hcal|}+ \frac{4f^2}{|\Hcal|^2}\Big)L^2\alpha_{k}^2\left\|\xbar_{k} - x_{\Hcal}^{\star}\right\|^2,\label{subsec:deterministic:T=1:eq2}
\end{align}
where the last inequality we use $|\Bcal_{k}| \leq f$. Using \eqref{thm:deterministic:T=1:f_cond} gives
\begin{align*}
\mu - \frac{2Lf}{|\Hcal|}  \geq \mu - \frac{2\mu}{3} = \frac{\mu}{3},  
\end{align*}
which when substituting into \eqref{subsec:deterministic:T=1:eq2} and using \eqref{thm:deterministic:T=1:stepsize} gives \eqref{thm:deterministic:T=1:ineq}
\begin{align*}
&\|\xbar_{k+1} - x_{\Hcal}^{\star}\|^2\notag\\ 
&\leq \left(1-\frac{2\mu}{3}\alpha_{k} \right)\left\|\xbar_{k} - x_{\Hcal}^{\star}\right\|^2\notag\\ 
&\quad + \Big(1 +\frac{4f}{|\Hcal|}+ \frac{4f^2}{|\Hcal|^2}\Big)L^2\alpha_{k}^2\left\|\xbar_{k} - x_{\Hcal}^{\star}\right\|^2\notag\\
&\leq \left(1-\frac{2\mu}{3}\alpha_{k} \right)\left\|\xbar_{k} - x_{\Hcal}^{\star}\right\|^2 + 3L^2\alpha_{k}^2\left\|\xbar_{k} - x_{\Hcal}^{\star}\right\|^2\notag\\
&\leq \left(1-\frac{\mu\alpha}{6} \right)\left\|\xbar_{k} - x_{\Hcal}^{\star}\right\|^2\notag\\
&\leq \left(1-\frac{\mu\alpha}{6} \right)^{k+1}\left\|\xbar_{0} - x_{\Hcal}^{\star}\right\|^2,
\end{align*}
where the second inequality is due to $f/|\Hcal| \leq 1/3$ and the third inequality is due to $\alpha_{k} = \alpha \leq \mu/(4L^2)$ 
\end{proof}
\subsection{Proof of Theorem~\ref{thm:deterministic:T>1}}
\label{app:deterministic:T>1}

\begin{proof}
Due to $\Tcal>1$, the proof of Theorem \ref{thm:deterministic:T>1} is different to the one in Theorem \ref{thm:deterministic:T=1} at the way we quantify the size of $\|x^{i}_{k,t} - \xbar_{k}\|$ for any $t\in[0,\Tcal]$. Thus, to show \eqref{thm:deterministic:T>1:ineq} we first provide an upper bound of this quantity. Note that we do not assume the gradient being bounded. By \eqref{thm:deterministic:T>1:stepsize} we have
\begin{align*}
L\Tcal\alpha_{k} =  L\Tcal\alpha\leq \frac{\mu}{16L} \leq \frac{1}{16}\leq \ln(2).      
\end{align*}
By \eqref{alg:x_i}, Assumption \ref{asp:lipschitz}, and $\nabla q^{i}(x_{\Hcal}^{\star}) = 0$ we have for all $t\in[0,\Tcal-1]$ and $i\in\Hcal$
\begin{align*}
&\|x_{k,t+1}^{i} - x_{\Hcal}^{\star}\| - \|x_{\Hcal}^{\star} - x_{k,t}^{i}\|\leq \|x_{k,t+1}^{i} - x_{k,t}^{i}\|  \notag\\  
&= \alpha_{k}\|\nabla q^{i}(x_{k,t}^{i})\|= \alpha_{k}\|\nabla q^{i}(x_{k,t}^{i}) - \nabla q^{i}(x_{\Hcal}^{\star})\| \notag\\ 
&\leq L\alpha_{k}\|x_{k,t}^{i} - x_{\Hcal}^{\star}\|,     
\end{align*}
which using $1+x\leq \exp(x)$ for $x>0$ implies
\begin{align*}
&\|x_{k,t+1}^{i} - x_{\Hcal}^{\star}\| \leq (1+L\alpha_{k})\|x_{k,t}^{i}- x_{\Hcal}^{\star}\|\notag\\
&\leq \exp(L\alpha_{k})\|x_{k,t}^{i}- x_{\Hcal}^{\star}\|\leq \exp(Lt\alpha_{k})\|x_{k,0}^{i}- x_{\Hcal}^{\star}\| \notag\\
&\leq 2\|\xbar_{k} - x_{\Hcal}^{\star}\|,
\end{align*}
where the last inequality we use $L\Tcal\alpha_{k} \leq \ln(2)$ and $x_{k,0}^{i} = \xbar_{k}$. Using the preceding two relations gives for all $t\in[0,\Tcal]$
\begin{align*}
&\|x_{k,t+1}^{i} - x_{k,t}^{i}\| \leq   L\alpha_{k}\|x_{k,t}^{i} - x_{\Hcal}^{\star}\| 
\leq 2L\alpha_{k}\|\xbar_{k} - x_{\Hcal}^{\star}\|.
\end{align*}
Using this relation and $x_{k,0}^{i} = \xbar_{k}$ gives $\forall t\in[0,\Tcal-1]$
\begin{align}
&\|x_{k,t+1}^{i} - \xbar_{k}\| = \Big\|\sum_{\ell = 0}^{t}x_{k,\ell+1}^{i} - x_{k,\ell}^{i}\Big\|\notag\\ 
&\leq    \sum_{\ell = 0}^{t}\Big\|x_{k,\ell+1}^{i} - x_{k,\ell}^{i}\Big\| \leq 2L\Tcal\alpha_{k}\|\xbar_{k}-x_{\Hcal}^{\star}\|. \label{subsec:deterministic:T>1:xi-xbar} 
\end{align}
Next, we consider 
\begin{align}
&\xbar_{k+1} = \frac{1}{|\Fcal_{k}|}\sum_{i\in\Fcal_{k}}x^{i}_{k,\Tcal}\notag\\ 
&=  \frac{1}{|\Hcal|}\Big[\sum_{i\in\Hcal}x^{i}_{k,\Tcal} + \sum_{i\in\Bcal_{k}}x^{i}_{k,\Tcal} - \sum_{i\in\Hcal\setminus\Hcal_{k}}x^{i}_{k,\Tcal}\Big]\notag\\  
&\stackrel{\eqref{subsec:deterministic:T>1:xi}}{=} \xbar_{k} - \frac{\alpha_{k}}{|\Hcal|}\sum_{i\in\Hcal}\sum_{t = 0}^{\Tcal - 1}\nabla q^{i}(x^{i}_{k,t})\notag\\  
&\quad + \frac{1}{|\Hcal|}\Big[ \sum_{i\in\Bcal_{k}}x^{i}_{k,\Tcal} - \sum_{i\in\Hcal\setminus\Hcal_{k}}x^{i}_{k,\Tcal}\Big]\notag\\
&= \xbar_{k} - \Tcal\alpha_{k}\nabla q^{\Hcal}(\xbar_{k})\notag\\
&\quad - \frac{\alpha_{k}}{|\Hcal|}\sum_{i\in\Hcal}\sum_{t = 0}^{\Tcal - 1}\left(\nabla q^{i}(x^{i}_{k,t}) - \nabla q^{i}(\xbar_{k})\right) \notag\\
&\quad  + \frac{1}{|\Hcal|}\Big[ \sum_{i\in\Bcal_{k}}x^{i}_{k,\Tcal} - \sum_{i\in\Hcal\setminus\Hcal_{k}}x^{i}_{k,\Tcal}\Big]\notag\\
&= \xbar_{k} - \Tcal\alpha_{k}\nabla q^{\Hcal}(\xbar_{k})\notag\\ 
&\quad - \frac{\alpha_{k}}{|\Hcal|}\sum_{i\in\Hcal}\sum_{t = 0}^{\Tcal - 1}\left(\nabla q^{i}(x^{i}_{k,t}) - \nabla q^{i}(\xbar_{k})\right) \notag\\
&\quad  + \frac{1}{|\Hcal|}\Big[ \sum_{i\in\Bcal_{k}}\!(x^{i}_{k,\Tcal} - \xbar_{k}) -\!\!\! \sum_{i\in\Hcal\setminus\Hcal_{k}}\!\!(x^{i}_{k,\Tcal}-\xbar_{k})\Big],\label{subsec:deterministic:T>1:xbar}
\end{align}
where the last equality is due to $|\Bcal_{k}| = |\Hcal\setminus\Hcal_{k}|$. For convenience, we denote by 
\begin{align*}
V_{k}^{x} = \frac{1}{|\Hcal|}\sum_{i\in\Bcal_{k}}\!(x^{i}_{k,\Tcal} - \xbar_{k}) -\!\!\! \sum_{i\in\Hcal\setminus\Hcal_{k}}\!\!(x^{i}_{k,\Tcal}-\xbar_{k}).     
\end{align*}
Using \eqref{subsec:deterministic:T>1:xbar} gives
\begin{align}
&\left\|\xbar_{k+1} - x_{\Hcal}^{\star}\right\|^2\notag\\ 
&= \big\|\xbar_{k} - x_{\Hcal}^{\star} - \Tcal\alpha_{k}\nabla q^{\Hcal}(\xbar_{k})\big\|^2 + \|V_{k}^{x}\|^2\notag\\ 
&\quad + \Big\|\frac{\alpha_{k}}{|\Hcal|}\sum_{i\in\Hcal}\sum_{t = 0}^{\Tcal - 1}\left(\nabla q^{i}(x^{i}_{k,t}) - \nabla q^{i}(\xbar_{k})\right)\Big\|^2\notag\\
&\quad - \frac{2\alpha_{k}}{|\Hcal|}\sum_{i\in\Hcal}\sum_{t = 0}^{\Tcal - 1}(\xbar_{k} - x_{\Hcal}^{\star})^T(\nabla q^{i}(x^{i}_{k,t}) - \nabla q^{i}(\xbar_{k}))\notag\\
&\quad + \frac{2\Tcal\alpha_{k}^2}{|\Hcal|} \sum_{i\in\Hcal}\sum_{t = 0}^{\Tcal - 1}\nabla q^{\Hcal}(\xbar_{k})^T\left(\nabla q^{i}(x^{i}_{k,t}) - \nabla q^{i}(\xbar_{k})\right)\notag\\
&\quad + 2\big(\xbar_{k} - x_{\Hcal}^{\star} - \Tcal\alpha_{k}\nabla q^{\Hcal}(\xbar_{k}) \big)^TV_{k}^{x}.\label{subsec:deterministic:T>1:xbar-x*}
\end{align}
We next consider each term on the right-hand sides of \eqref{subsec:deterministic:T>1:xbar-x*}. First, using Assumptions \ref{asp:lipschitz} and \ref{asp:str_cvxty}, and $\nabla q^{\Hcal}(x^{\star}_{\Hcal}) = 0$ we have
\begin{align}
&\Big\|\xbar_{k} - x_{\Hcal}^{\star} - \Tcal\alpha_{k}\nabla q^{\Hcal}(\xbar_{k})\Big\|^2\notag\\ 
&= \Big\|\xbar_{k} - x_{\Hcal}^{\star}\Big\|^2  -  2\Tcal\alpha_{k}\left(\xbar_{k} - x_{\Hcal}^{\star}\right)^T\nabla q^{\Hcal}(\xbar_{k})\notag\\ 
&\quad +  \Tcal^2\alpha_{k}^2\left\|\nabla q^{\Hcal}(\xbar_{k})\right\|^2\notag\\
&= \Big\|\xbar_{k} - x_{\Hcal}^{\star}\Big\|^2  -  2\Tcal\alpha_{k}\left(\xbar_{k} - x_{\Hcal}^{\star}\right)^T\nabla q^{\Hcal}(\xbar_{k})\notag\\ 
&\quad + \Tcal^2\alpha_{k}^2\left\| \nabla q^{\Hcal}(\xbar_{k})-\nabla q^{\Hcal}(x_{\Hcal}^{\star})\right\|^2\notag\\
&\leq \big(1 - 2\mu \Tcal\alpha_{k} + \Tcal^2L^2\alpha_{k}^2\big)\left\|\xbar_{k} - x_{\Hcal}^{\star}\right\|^2.\label{subsec:deterministic:T>1:Eq1a}
\end{align}
Second, by \eqref{alg:CEfilter} there exists $j\in\Hcal\setminus\Hcal_{k}$ such that $\|x^{i}_{k,\Tcal} - \xbar_{k}\| \leq \|x^{j}_{k,\Tcal} - \xbar_{k}\|$ for all $i\in\Bcal_{k}$. Then we have
\begin{align}
&\|V_{k}^{x}\|^2 = \Big\|\frac{1}{|\Hcal|}\!\! \sum_{i\in\Bcal_{k}}\!x^{i}_{k,\Tcal} - \xbar_{k} - \frac{1}{|\Hcal|}\!\! \sum_{i\in\Hcal\setminus\Hcal_{k}}\!\!\!x^{i}_{k,\Tcal}-\xbar_{k}\Big\|^2\notag\\
&\leq \frac{2|\Bcal_{k}|}{|\Hcal|^2}\!\!\sum_{i\in\Bcal_{k}}\!\!\left\|x^{i}_{k,\Tcal} - \xbar_{k}\right\|^2 + \frac{2|\Bcal_{k}|}{|\Hcal|^2}\!\!\!\sum_{i\in\Hcal\setminus\Hcal_{k}}\!\!\!\left\|x^{i}_{k,\Tcal}-\xbar_{k}\right\|^2\notag\\
&\leq \frac{2|\Bcal_{k}|^2}{|\Hcal|^2}\left\|x^{j}_{k,\Tcal} - \xbar_{k}\right\|^2 + \frac{2|\Bcal_{k}|}{|\Hcal|^2}\sum_{i\in\Hcal\setminus\Hcal_{k}}\left\|x^{i}_{k,\Tcal}-\xbar_{k}\right\|^2\notag\\
&\stackrel{\eqref{subsec:deterministic:T>1:xi-xbar}}{\leq} \frac{16\Tcal^2L^2|\Bcal_{k}|^2}{|\Hcal|^2} \alpha_{k}^2\left\|\xbar_{k} - x_{\Hcal}^{\star}\right\|^2. \label{subsec:deterministic:T>1:Eq1b}  
\end{align}
Third, using \eqref{subsec:deterministic:T>1:xi-xbar} yields
\begin{align}
&\Big\|\frac{\alpha_{k}}{|\Hcal|}\sum_{i\in\Hcal}\sum_{t = 0}^{\Tcal - 1}\left(\nabla q^{i}(x^{i}_{k,t}) - \nabla q^{i}(\xbar_{k})\right)\Big\|^2\notag\\
&\leq \frac{L^2\Tcal\alpha_{k}^2}{|\Hcal|}\sum_{i\in\Hcal}\sum_{t = 0}^{\Tcal - 1}\Big\|x^{i}_{k,t} - \xbar_{k}\Big\|^2\notag\\
&\leq 4L^4\Tcal^{4}\alpha_{k}^4\Big\|\xbar_{k} - x_{\Hcal}^{\star}\Big\|^2.\label{subsec:deterministic:T>1:Eq1c}  
\end{align}
Fourth, using Assumption \ref{asp:lipschitz} we consider 
\begin{align}
&-\frac{2\alpha_{k}}{|\Hcal|}\Big(\xbar_{k} - x_{\Hcal}^{\star}\Big)^T   \sum_{i\in\Hcal}\sum_{t = 0}^{\Tcal - 1}\left(\nabla q^{i}(x^{i}_{k,t}) - \nabla q^{i}(\xbar_{k})\right)\notag\\
&\leq \frac{2L\alpha_{k}}{|\Hcal|}   \sum_{i\in\Hcal}\sum_{t = 0}^{\Tcal - 1}\|\xbar_{k} - x_{\Hcal}^{\star}\| \|x^{i}_{k,t} - \xbar_{k}\|\notag\\
&\stackrel{\eqref{subsec:deterministic:T>1:xi-xbar}}{\leq} 4\Tcal^2L^2\alpha_{k}^2\|\xbar_{k}-x_{\Hcal}^{\star}\|^2.  \label{subsec:deterministic:T>1:Eq1d}
\end{align}
Fifth, using Assumption \ref{asp:lipschitz} and $\nabla q^{\Hcal}(x_{\Hcal}^{\star}) = 0$ we have
\begin{align}
&\frac{2\Tcal\alpha_{k}^2}{|\Hcal|} \sum_{i=1}^{|\Hcal|}\sum_{t = 0}^{\Tcal - 1}\nabla q^{\Hcal}(\xbar_{k})^T\left(\nabla q^{i}(x^{i}_{k,t}) - \nabla q^{i}(\xbar_{k})\right)\notag\\   
&\leq \frac{2L\Tcal\alpha_{k}^2}{|\Hcal|} \sum_{i=1}^{|\Hcal|}\sum_{t = 0}^{\Tcal - 1}\|\nabla q^{\Hcal}(\xbar_{k})\|\|x^{i}_{k,t} - \xbar_{k}\|\notag\\
&\stackrel{}{\leq}4L^2\Tcal^{3}\alpha_{k}^3\|\nabla q^{\Hcal}(\xbar_{k})\|\|\xbar_{k} - x_{\Hcal}^{\star}\|\notag\\
&= 4L^2\Tcal^{3}\alpha_{k}^3 \|\nabla q^{\Hcal}(\xbar_{k})-\nabla q^{\Hcal}(x_{\Hcal}^{\star})\|\|\xbar_{k} - x_{\Hcal}^{\star}\|\notag\\  
&\leq 4L^3\Tcal^{3}\alpha_{k}^3\|\xbar_{k} - x_{\Hcal}^{\star}\|^2.\label{subsec:deterministic:T>1:Eq1e}
\end{align}
Next, by using \eqref{alg:CEfilter} and similar to \eqref{subsec:deterministic:T>1:Eq1b} we have 
\begin{align*}
&\|V_{k}^{x}\| = \Big\|\frac{1}{|\Hcal|}\!\! \sum_{i\in\Bcal_{k}}\!x^{i}_{k,\Tcal} - \xbar_{k} - \frac{1}{|\Hcal|}\!\! \sum_{i\in\Hcal\setminus\Hcal_{k}}\!\!\!x^{i}_{k,\Tcal}-\xbar_{k}\Big\|\notag\\
&\leq \frac{1}{|\Hcal|}\!\!\sum_{i\in\Bcal_{k}}\!\!\left\|x^{i}_{k,\Tcal} - \xbar_{k}\right\| + \frac{1}{|\Hcal|}\!\!\!\sum_{i\in\Hcal\setminus\Hcal_{k}}\!\!\!\left\|x^{i}_{k,\Tcal}-\xbar_{k}\right\|\notag\\
&\stackrel{\eqref{subsec:deterministic:T>1:xi-xbar}}{\leq} \frac{2\Tcal L|\Bcal_{k}|}{|\Hcal|} \alpha_{k}\left\|\xbar_{k} - x_{\Hcal}^{\star}\right\|,    
\end{align*}
which by using $\nabla q^{\Hcal}(x_{\Hcal}^{\star}) = 0$, \eqref{alg:CEfilter}, and \eqref{subsec:deterministic:T>1:Eq1c} we have  
\begin{align}
&2\Big(\xbar_{k} - x_{\Hcal}^{\star} - \Tcal\alpha_{k}\nabla q^{\Hcal}(\xbar_{k}) \Big)^TV_{k}^{x}\notag\\
&\leq 2\left(\left\|\xbar_{k} - x_{\Hcal}^{\star}\right\| + \Tcal\alpha_{k}\left\|\nabla q^{\Hcal}(\xbar_{k}) - \nabla g(\xbar^{\star})\right \|\right)\|V_{k}^{x}\|\notag\\
&\stackrel{\eqref{subsec:deterministic:T>1:Eq1b}}{\leq} \frac{4\Tcal L|\Bcal_{k}|}{|\Hcal|}(1+L\Tcal\alpha_{k})\alpha_{k}\left\|\xbar_{k} - x_{\Hcal}^{\star}\right\|^2. \label{subsec:deterministic:T>1:Eq1f}
\end{align}
Substituting \eqref{subsec:deterministic:T>1:Eq1a}--\eqref{subsec:deterministic:T>1:Eq1f} into \eqref{subsec:deterministic:T>1:xbar-x*} and using $|\Bcal_{k}| \leq f$ yields 
\begin{align}
&\left\|\xbar_{k+1} - x_{\Hcal}^{\star}\right\|^2\notag\\ 
&\leq  \big(1 - 2\mu \Tcal\alpha_{k} + \Tcal^2L^2\alpha_{k}^2\big)\left\|\xbar_{k} - x_{\Hcal}^{\star}\right\|^2\notag\\
&\quad + \frac{16\Tcal^2L^2f^2}{|\Hcal|^2} \alpha_{k}^2\left\|\xbar_{k} - x_{\Hcal}^{\star}\right\|^2 + 4L^4\Tcal^{4}\alpha_{k}^4\Big\|\xbar_{k} - x_{\Hcal}^{\star}\Big\|^2\notag\\
&\quad + 4L^2\Tcal^{2}\alpha_{k}^2\Big\|\xbar_{k} - x_{\Hcal}^{\star}\Big\|^2 + 4L^3\Tcal^{3}\alpha_{k}^3\|\xbar_{k} - x_{\Hcal}^{\star}\|^2\notag\\
&\quad + \frac{4\Tcal Lf}{|\Hcal|}(1+L\Tcal\alpha_{k})\alpha_{k}\left\|\xbar_{k} - x_{\Hcal}^{\star}\right\|^2\notag\\
&=  \Big(1 - 2\Tcal\big(\mu - \frac{2Lf}{|\Hcal|}\big)\alpha_{k}\Big)\left\|\xbar_{k} - x_{\Hcal}^{\star}\right\|^2\notag\\
&\quad +  4L^4\Tcal^{4}\alpha_{k}^4\Big\|\xbar_{k} - x_{\Hcal}^{\star}\Big\|^2 + 4L^3\Tcal^{3}\alpha_{k}^3\|\xbar_{k} - x_{\Hcal}^{\star}\|^2\notag\\
&\quad + \Big(5 + \frac{4f}{|\Hcal|} + \frac{16f^2}{|\Hcal|^2}\Big)\Tcal^2L^2\alpha_{k}^2\left\|\xbar_{k} - x_{\Hcal}^{\star}\right\|^2.\label{subsec:deterministic:T>1:Eq1}
\end{align}
By \eqref{thm:deterministic:T=1:f_cond} we have
\begin{align*}
\mu - \frac{2Lf}{|\Hcal|} \geq \frac{\mu}{3}.     
\end{align*}
Moreover, using $f/|\Hcal| \leq \frac{1}{3}$ and $\alpha_{k} = \alpha \leq 1/(16\Tcal L)$ (since $\mu/L\leq1$) yields
\begin{align*}
\frac{16f^2}{|\Hcal|^2} + 4L^2\Tcal^2\alpha_{k}^2 + 5 + 4L\Tcal\alpha_{k} + \frac{4 f}{|\Hcal|} \leq  8.  
\end{align*}
Using the preceding two relations into \eqref{thm:deterministic:T>1:ineq} gives
\begin{align*}
&\left\|\xbar_{k+1} - x_{\Hcal}^{\star}\right\|^2\notag\\ 
&\leq  \Big(1 - \frac{2\mu \Tcal}{3}\alpha\Big)\left\|\xbar_{k} - x_{\Hcal}^{\star}\right\|^2 + 8L^2\Tcal^{2}\alpha^2\Big\|\xbar_{k} - x_{\Hcal}^{\star}\Big\|^2\notag\\
&\leq \Big(1 - \frac{\mu \Tcal \alpha}{6}\Big)\Big\|\xbar_{k} - x_{\Hcal}^{\star}\Big\|^2\notag\\
&\leq \Big(1 - \frac{\mu \Tcal \alpha}{6}\Big)^{k+1}\Big\|\xbar_{0} - x_{\Hcal}^{\star}\Big\|^2,
\end{align*}
where in the second inequality we use $\alpha_{k} = \alpha \leq \mu/(16 \Tcal L^2)$.
\end{proof}
\subsection{Proof of Theorem~\ref{thm:stochastic:T=1}}
\label{app:stochastic:T=1}

\begin{proof}
When $T=1$ and by \eqref{alg:x_i} with $G^{i}(\cdot) = \nabla Q^{i}(\cdot,X^{i})$ we have for all $i\in\Hcal$
\begin{align}
x^{i}_{k,1} &= x^{i}_{k,0} - \alpha_{k}\nabla Q^{i} (x^{i}_{k,0};X^{i}_{k,0})\notag\\ 
&= \xbar_{k} - \alpha_{k}\nabla Q^{i} (\xbar_{k};X^{i}_{k,0}),\label{subsec:stochastic:T=1:xi}     
\end{align}
which by \eqref{alg:xbar} gives
\begin{align*}
&\xbar_{k+1} = \frac{1}{|\Fcal_{k}|}\sum_{i\in\Fcal_{k}}x^{i}_{k,1}\notag\\ 
&=  \frac{1}{|\Hcal|}\Big[\sum_{i\in\Hcal}x^{i}_{k,1} + \sum_{i\in\Bcal_{k}}x^{i}_{k,1} - \sum_{i\in\Hcal\setminus\Hcal_{k}}x^{i}_{k,1}\Big]\notag\\  
&\stackrel{\eqref{subsec:stochastic:T=1:xi}}{=} \xbar_{k} - \alpha_{k}\nabla Q(\xbar_{k};X_{k,0})\notag\\ 
&\quad + \frac{1}{|\Hcal|}\Big[ \sum_{i\in\Bcal_{k}}x^{i}_{k,1} - \sum_{i\in\Hcal\setminus\Hcal_{k}}x^{i}_{k,1}\Big]\notag\\
&= \xbar_{k} - \alpha_{k}\nabla Q(\xbar_{k};X_{k,0})\notag\\ 
&\quad + \frac{1}{|\Hcal|}\Big[ \sum_{i\in\Bcal_{k}}x^{i}_{k,1} - \xbar_{k} - \sum_{i\in\Hcal\setminus\Hcal_{k}}x^{i}_{k,1}-\xbar_{k}\Big],
\end{align*}
where the last equality is due to $|\Bcal_{k}| = |\Hcal\setminus\Hcal_{k}|$. Using the preceding relation, we consider
\begin{align}
&\|\xbar_{k+1} - x_{\Hcal}^{\star}\|^2\notag\\ 
&= \left\|\xbar_{k} - x_{\Hcal}^{\star} - \alpha_{k}\nabla Q(\xbar_{k};X_{k,0})\right\|^2\notag\\ 
&\quad + \frac{1}{|\Hcal|^2}\Big\| \sum_{i\in\Bcal_{k}}(x^{i}_{k,1} - \xbar_{k}) - \sum_{i\in\Hcal\setminus\Hcal_{k}}(x^{i}_{k,1}-\xbar_{k}) \Big\|^2\notag\\
&\quad + \frac{2}{|\Hcal|}\sum_{i\in\Bcal_{k}}\left(\xbar_{k} - x_{\Hcal}^{\star}\right)^T(x^{i}_{k,1} - \xbar_{k})\notag\\
&\quad - \frac{2\alpha_{k}}{|\Hcal|}\sum_{i\in\Bcal_{k}}\nabla Q(\xbar_{k};X_{k,0})^T(x^{i}_{k,1} - \xbar_{k})\notag\\
&\quad - \frac{2}{|\Hcal|}\sum_{i\in\Hcal\setminus\Hcal_{k}}\left(\xbar_{k} - x_{\Hcal}^{\star} \right)^T(x^{i}_{k,1} - \xbar_{k})\notag\\
&\quad + \frac{2\alpha_{k}}{|\Hcal|}\sum_{i\in\Hcal\setminus\Hcal_{k}}\nabla Q(\xbar_{k};X_{k,0})^T(x^{i}_{k,1} - \xbar_{k}).\label{subsec:stochastic:T=1:eq1}
\end{align}
We next analyze each term on the right-hand side of \eqref{subsec:stochastic:T=1:eq1}. First, using Assumption \ref{assump:noise} we have
\begin{align*}
&\Eset[\left\|\xbar_{k} - x_{\Hcal}^{\star} - \alpha_{k}\nabla Q(\xbar_{k};X_{k,0})\right\|^2\,|\,\Pcal_{k,0}]\notag\\
&= \|\xbar_{k} - x_{\Hcal}^{\star} - \alpha_{k}\nabla q(\xbar_{k})\|^2 \notag\\
&\quad + \Eset[\|\alpha_{k}(\nabla Q(\xbar_{k};X_{k,0})-\nabla q(\xbar_{k}))\|^2\,|\,\Pcal_{k,0}]\notag\\
&\leq \|\xbar_{k} - x_{\Hcal}^{\star} - \alpha_{k}\nabla q(\xbar_{k})\|  + \sigma^2\alpha_{k}^2,
\end{align*}
which by using Assumption \ref{asp:str_cvxty}  and $\nabla q(x_{\Hcal}^{\star}) = 0$ gives
\begin{align}
&\Eset[\left\|\xbar_{k} - x_{\Hcal}^{\star} - \alpha_{k}\nabla Q(\xbar_{k};X_{k,0})\right\|^2\,|\,\Pcal_{k,0}]\notag\\
&=\|\xbar_{k} - x_{\Hcal}^{\star}\|^2 - 2 ( \nabla q(\xbar_{k})-\nabla q(x_{\Hcal}^{\star}))^T(\xbar_{k} - x_{\Hcal}^{\star}) \notag\\ 
&\quad + \alpha_{k}^2\|\nabla q(\xbar_{k}) - \nabla q(x_{\Hcal}^{\star})\|^2 + \sigma^2\alpha_{k} \notag\\
&\leq (1-2\mu\alpha_{k}+ L^2\alpha_{k}^2)\left\|\xbar_{k} - x_{\Hcal}^{\star}\right\|^2 + \sigma^2\alpha_{k}^2.   \label{subsec:stochastic:T=1:eq1a}
\end{align}
Second, by \eqref{eqn:int_min_sets} we have $\nabla q^{i}(x_{\Hcal}^{\star}) = 0$. Then, by Assumption \ref{asp:lipschitz} and \eqref{subsec:stochastic:T=1:xi} we have
\begin{align}
&-\frac{2}{|\Hcal|}  \Eset\Big[\sum_{i\in\Hcal\setminus\Hcal_{k}}(\xbar_{k} - x_{\Hcal}^{\star})^T(x^{i}_{k,1}-\xbar_{k})\;|\;\Pcal_{k,0}\Big]\notag\\ 
&= \frac{2\alpha_{k}}{|\Hcal|}  \sum_{i\in\Hcal\setminus\Hcal_{k}}(\xbar_{k} - x_{\Hcal}^{\star})^T\Eset\left[\nabla Q^{i}(\xbar_{k};X^{i}_{k,0})\;|\;\Pcal_{k,0}\right]\notag\\
&= \frac{2\alpha_{k}}{|\Hcal|}  \sum_{i\in\Hcal\setminus\Hcal_{k}}(\xbar_{k} - x_{\Hcal}^{\star})^T\nabla q^{i}(\xbar_{k})\notag\\ 
&= \frac{2\alpha_{k}}{|\Hcal|}  \sum_{i\in\Hcal\setminus\Hcal_{k}}(\xbar_{k} - x_{\Hcal}^{\star})^T(\nabla q^{i}(\xbar_{k}) - \nabla q^{i}(x_{\Hcal}^{\star}))\notag\\ 
&\leq \frac{2L|\Bcal_{k}|\alpha_{k}}{|\Hcal|}\|\xbar_{k}-x_{\Hcal}^{\star}\|^2,\label{subsec:stochastic:T=1:eq1b}
\end{align}
where the last inequality we also use $|\Hcal\setminus\Hcal_{k}| = |\Bcal_{k}|$. Next, using Assumption \ref{assump:noise} we consider for any $i\in\Hcal$
\begin{align*}
&\Eset\Big[\nabla Q(\xbar_{k};X_{k,0})^T\nabla Q^{i}(\xbar_{k};X^{i}_{k,0})\;|\;\Pcal_{k,0}\Big]\notag\\
&= \frac{1}{|\Hcal|}\Eset\Big[\sum_{j\in\Hcal}\nabla Q^{j}(\xbar_{k};X_{k,0}^{j})^T\nabla Q^{i}(\xbar_{k};X^{i}_{k,0})\;|\;\Pcal_{k,0}\Big]\notag\\
&= \frac{1}{|\Hcal|}\Eset[\|\nabla Q^{i}(\xbar_{k};X_{k,0}^{i})\|^2\;|\;\Pcal_{k,0}]\notag\\
&\quad+ \frac{1}{|\Hcal|}\sum_{j\in\Hcal,i\neq j}\nabla q^{j}(\xbar_{k})^T\nabla q^{i}(\xbar_{k})\notag\\
&= \frac{1}{|\Hcal|}\|\nabla q^{i}(\xbar_{k})\|^2+ \frac{1}{|\Hcal|}\sum_{j\in\Hcal,i\neq j}\nabla q^{j}(\xbar_{k})^T\nabla q^{i}(\xbar_{k})\notag\\
&\quad +\frac{1}{|\Hcal|}\Eset[\|\nabla Q^{i}(\xbar_{k};X_{k,0}^{i})-\nabla q^{i}(\xbar_{k})\|^2\;|\;\Pcal_{k,0}]\notag\\
&\leq \frac{\sigma^2}{|\Hcal|} + \frac{L^2}{|\Hcal|}\sum_{j\in\Hcal}\|\xbar_{k}-x_{\Hcal}^{\star}\|^2\notag\\
&\leq \frac{\sigma^2}{|\Hcal|} + L^2\|\xbar_{k}-x_{\Hcal}^{\star}\|^2,
\end{align*}
where the second inequality we use $\nabla q^{i}(x_{\Hcal}^{\star}) = 0$ for all $i$ and Assumption \ref{asp:lipschitz}. Using the preceding relation we have
\begin{align}
&-\frac{2\alpha_{k}}{|\Hcal|}  \Eset\Big[\sum_{i\in\Bcal_{k}}\nabla Q(\xbar_{k};X_{k,0})^T(x^{i}_{k,1}-\xbar_{k})\;|\;\Pcal_{k,0}\Big]\notag\\  
&= \frac{2\alpha_{k}^2}{|\Hcal|}  \Eset\Big[\sum_{i\in\Bcal_{k}}\nabla Q(\xbar_{k};X_{k,0})^T\nabla Q^{i}(\xbar_{k};X^{i}_{k,0})\;|\;\Pcal_{k,0}\Big]\notag\\  
&\leq \frac{2\sigma^2|\Bcal_{k}|\alpha_{k}^2}{|\Hcal|^2} + \frac{2L^2|\Bcal_{k}|\alpha_{k}^2}{|\Hcal|}\|\xbar_{k}-x_{\Hcal}^{\star}\|^2.   \label{subsec:stochastic:T=1:eq1c}
\end{align}
Using the same argument as above, we have
\begin{align}
&\frac{2\alpha_{k}}{|\Hcal|}\Eset\Big[\sum_{i\in\Hcal\setminus\Hcal_{k}}\nabla Q(\xbar_{k};X_{k,0})^T(x^{i}_{k,1} - \xbar_{k})\,|\,\Pcal_{k,0}\Big]\notag\\
&\leq \frac{2\sigma^2|\Bcal_{k}|\alpha_{k}^2}{|\Hcal|^2} + \frac{2L^2|\Bcal_{k}|\alpha_{k}^2}{|\Hcal|}\|\xbar_{k}-x_{\Hcal}^{\star}\|^2, \label{subsec:stochastic:T=1:eq1d}
\end{align}
where we use the fact that $|\Bcal_{k}| = |\Hcal\setminus\Hcal_{k}|$. Fifth, using \eqref{subsec:stochastic:T=1:xi} and our CE filter \eqref{alg:CEfilter} there exists $j\in\Hcal$ such that for all $i\in\Bcal_{k}$ we have
\begin{align*}
&\|x^{i}_{k,1}-\xbar_{k}\| \leq \|x^{j}_{k,1}-\xbar_{k}\| = \alpha_{k}\|\nabla Q^{j}(\xbar_{k};X^{j}_{k,0})\|\notag\\
&\leq \alpha_{k}\|\nabla q^{j}(\xbar_{k})\| + \alpha_{k}\|\nabla Q^{j}(\xbar_{k};X^{j}_{k,0}) -\nabla q^{j}(\xbar_{k})\|,   
\end{align*}
which by Assumption \ref{assump:noise} and the Jensen inequality gives
\begin{align*}
\Eset\left[\|x^{i}_{k,1}-\xbar_{k}\|\,|\,\Pcal_{k,0}\right] \leq L\alpha_{k}\|\xbar_{k}-x_{\Hcal}^{\star}\| +  \sigma\alpha_{k}.    
\end{align*}
Similarly, for all $i\in\Bcal_{k}$ there exists $j\in\Hcal$ such that
\begin{align*}
&\Eset[\|x^{i}_{k,1}-\xbar_{k}\|^2\,|\,\Pcal_{k,0}]\notag\\
&\leq \alpha_{k}^2\|\nabla q^{j}(\xbar_{k})\|^2\notag\\
&\quad + \alpha_{k}^2\Eset[\|\nabla Q^{j}(\xbar_{k};X^{j}_{k,0}) -\nabla q^{j}(\xbar_{k})\|^2 \,|\,\Pcal_{k,0}]\notag\\
&\leq L^2\alpha_{k}^2\|\xbar_{k}-x_{\Hcal}^{\star}\|^2 + \sigma^2\alpha_{k}^2.
\end{align*}
Using the preceding relation we consider 
\begin{align}
&\frac{2}{|\Hcal|}\Eset\Big[\sum_{i\in\Bcal_{k}}\left(\xbar_{k} - x_{\Hcal}^{\star}\right)^T(x^{i}_{k,1} - \xbar_{k})\,|\,\Pcal_{k,0}\Big]\notag\\ 
&\leq \frac{2L\alpha_{k}}{|\Hcal|}\sum_{i\in\Bcal_{k}}\|\xbar_{k} - x_{\Hcal}^{\star}\|^2 +  \frac{2\sigma\alpha_{k}}{|\Hcal|}\sum_{i\in\Bcal_{k}}\|\xbar_{k} - x_{\Hcal}^{\star}\| \notag\\
&\leq \frac{3L|\Bcal_{k}|\alpha_{k}}{|\Hcal|}\|\xbar_{k}-x_{\Hcal}^{\star}\|^2 + \frac{\sigma^2|\Bcal_{k}|\alpha_{k}}{L|\Hcal|},  \label{subsec:stochastic:T=1:eq1dd}
\end{align}
where the last inequality we use the relation $2xy \leq \eta x^2 + 1/\eta y^2$ for any $x,y\in\Rset$. Finally, we have
\begin{align}
&\frac{1}{|\Hcal|^2}\Eset\Big[\big\| \sum_{i\in\Bcal_{k}}x^{i}_{k,1} - \xbar_{k} - \sum_{i\in\Hcal\setminus\Hcal_{k}}\!\!x^{i}_{k,1}-\xbar_{k} \big\|^2\,|\,\Pcal_{k,0}\Big]\notag\\  
&\leq \frac{2|\Bcal_{k}|}{|\Hcal|^2}\sum_{i\in\Bcal_{k}}\Eset\Big[\|x^{i}_{k,1} - \xbar_{k}\|^2\,|\,\Pcal_{k,0}\Big]\notag\\
&\quad +\frac{2|\Bcal_{k}|}{|\Hcal|^2} \sum_{i\in\Hcal\setminus\Hcal_{k}}\Eset\Big[\|x^{i}_{k,1}-\xbar_{k}\|^2 \,|\,\Pcal_{k,0}\Big]\notag\\
&\leq \frac{2|\Bcal_{k}|^2}{|\Hcal|^2}\Big(\sigma^2\alpha_{k}^2 + 2L^2\alpha_{k}^2\|\xbar_{k}-x_{\Hcal}^{\star}\|^2\Big).
\label{subsec:stochastic:T=1:eq1e}
\end{align}
Taking the expectation on both sides of \eqref{subsec:stochastic:T=1:eq1} and using  \eqref{subsec:stochastic:T=1:eq1a}--\eqref{subsec:stochastic:T=1:eq1e}  we obtain
\begin{align*}
&\Eset[\|\xbar_{k+1} - x_{\Hcal}^{\star}\|^2]\notag\\ 
&\leq (1-2\mu\alpha_{k}+L^2\alpha_{k}^2)\Eset[\|\xbar_{k} - x_{\Hcal}^{\star}\|^2] + \sigma^2\alpha_{k}^2\notag\\
&\quad + \frac{2L|\Bcal_{k}|\alpha_{k}}{|\Hcal|}\Eset[\|\xbar_{k}-x_{\Hcal}^{\star}\|^2]\notag\\
&\quad + \frac{2\sigma^2|\Bcal_{k}|\alpha_{k}^2}{|\Hcal|^2} + \frac{2L^2|\Bcal_{k}|\alpha_{k}^2}{|\Hcal|}\Eset[\|\xbar_{k}-x_{\Hcal}^{\star}\|^2]\notag\\
&\quad + \frac{2\sigma^2|\Bcal_{k}|\alpha_{k}^2}{|\Hcal|^2} + \frac{2L^2|\Bcal_{k}|\alpha_{k}^2}{|\Hcal|}\Eset[\|\xbar_{k}-x_{\Hcal}^{\star}\|^2]\notag\\
&\quad + \frac{3L|\Bcal_{k}|\alpha_{k}}{|\Hcal|}\Eset[\|\xbar_{k}-x_{\Hcal}^{\star}\|^2] + \frac{\sigma^2|\Bcal_{k}|\alpha_{k}}{L|\Hcal|}\notag\\
&\quad + \frac{2\sigma^2|\Bcal_{k}|^2\alpha_{k}^2}{|\Hcal|^2} + \frac{4L^2|\Bcal_{k}|^2\alpha_{k}^2}{|\Hcal|^2}\Eset[\|\xbar_{k}-x_{\Hcal}^{\star}\|^2]\notag\\
&\leq \Big(1-2\big(\mu-\frac{5Lf}{2|\Hcal|}\big)\alpha_{k}\Big)\Eset[\|\xbar_{k} - x_{\Hcal}^{\star}\|^2] + 7\sigma^2\alpha_{k}^2 \notag\\
&\quad + L^2\alpha_{k}^2\Eset[\|\xbar_{k} - x_{\Hcal}^{\star}\|^2]   +  \frac{4L^2f\alpha_{k}^2}{|\Hcal|}\Eset[\|\xbar_{k}-x_{\Hcal}^{\star}\|^2]\notag\\
&\quad + \frac{4L^2f^2\alpha_{k}^2}{|\Hcal|^2}\Eset[\|\xbar_{k}-x_{\Hcal}^{\star}\|^2]+ \frac{\sigma^2f}{L|\Hcal|}\alpha_{k},
\end{align*}
where we use $|\Bcal_{k}| = f \leq |\Hcal|$ in the last inequality. Note that by \eqref{thm:deterministic:T=1:f_cond} we have $3Lf/|\Hcal| \leq \mu$ and $f/|\Hcal| \leq 1/3$, which when substituting to the preceding equation gives
\begin{align*}
&\Eset[\|\xbar_{k+1} - x_{\Hcal}^{\star}\|^2]\notag\\   
&\leq \left(1-\frac{\mu}{3}\alpha_{k}\right)\Eset[\|\xbar_{k} - x_{\Hcal}^{\star}\|^2] + 7\sigma^2\alpha_{k}^2 + \frac{\sigma^2f}{L|\Hcal|}\alpha_{k}\notag\\
&\quad +  4L^2\alpha_{k}^2\Eset[\|\xbar_{k}-x_{\Hcal}^{\star}\|^2]\notag\\
&\leq \left(1-\frac{\mu}{6}\alpha\right)\Eset[\|\xbar_{k} - x_{\Hcal}^{\star}\|^2] + 7\sigma^2\alpha^2 + \frac{\sigma^2f}{L|\Hcal|}\alpha,
\end{align*}
where the last inequality we use $\alpha_{k} = \alpha \leq \mu/(12L^2)$. Taking recursively the previous inequality immediately gives \eqref{thm:stochastic:T=1:ineq}.
% \begin{align*}
% &\Eset[\|\xbar_{k+1} - x_{\Hcal}^{\star}\|^2]\notag\\ 
% &\leq \left(1-\frac{\mu}{2}\alpha\right)^{k+1}\Eset[\|\xbar_{0} - x_{\Hcal}^{\star}\|^2] + \frac{14\sigma^2\alpha}{\mu} + \frac{2\sigma^2f}{\mu L|\Hcal|}\cdot
% \end{align*}
\end{proof}

\subsection{Proof of Theorem \ref{thm:stochastic:T>1}}
\begin{proof}
Using \eqref{alg:x_i} we have for all $ t\in[0,\Tcal-1]$ and $i\in\Hcal$
\begin{align*}
&\|x_{k,t+1}^{i} - x_{\Hcal}^{\star}\| - \|x_{\Hcal}^{\star} - x_{k,t}^{i}\|\notag\\
&\leq \|x_{k,t+1}^{i} - x_{k,t}^{i}\|  = \alpha_{k}\|\nabla Q^{i}(x_{k,t}^{i};X_{k,t}^{i})\|\notag\\
&\leq \alpha_{k}\|\nabla q^{i}(x_{k,t}^{i})\|  + \alpha_{k}\|\nabla Q^{i}(x_{k,t}^{i};X_{k,t}^{i})-\nabla q^{i}(x_{k,t}^{i})\|,
\end{align*}
which by using Assumptions \ref{asp:lipschitz} and \ref{assump:noise}, and $\nabla q^{i}(x_{\Hcal}^{\star}) = 0$ yields 
\begin{align*}
&\Eset[\|x_{k,t+1}^{i} - x_{\Hcal}^{\star}\| - \|x_{\Hcal}^{\star} - x_{k,t}^{i}\|\,|\,\Pcal_{k,t}]\notag\\
&\leq\Eset[\|x_{k,t+1}^{i}  - x_{k,t}^{i}\|\,|\,\Pcal_{k,t}]\notag\\
&\leq \alpha_{k}\|\nabla q^{i}(x_{k,t}^{i})-\nabla q^{i}(x_{\Hcal}^{\star})\|\notag\\ 
&\quad + \alpha_{k}\Eset[\|\nabla Q^{i}(x_{k,t}^{i};X_{k,t}^{i})-\nabla q^{i}(x_{k,t}^{i})\|\,|\,\Pcal_{k,t}]\notag\\
&\leq L\alpha_{k}\|x_{k,t}^{i} - x_{\Hcal}^{\star}\| + \sigma\alpha_{k}.
\end{align*}
Using $1+x\leq \exp(x)$ for $x>0$, the preceding relation gives
\begin{align*}
&\Eset[\|x_{k,t+1}^{i} - x_{\Hcal}^{\star}\|] \leq (1+L\alpha_{k})\Eset[\|x_{k,t}^{i}- x_{\Hcal}^{\star}\|] + \sigma \alpha_{k}\notag\\
&\leq (1+L\alpha_{k})^{t}\Eset[\|x_{k,0}^{i}- x_{\Hcal}^{\star}\|] + \sigma \alpha_{k}\sum_{\ell=0}^{t-1}(1+L\alpha_{k})^{t-1-\ell}\notag\\
&= \exp(Lt\alpha_{k})\Eset[\|x_{k,0}^{i}- x_{\Hcal}^{\star}\|]\notag\\ 
&\quad + \sigma \alpha_{k}(1+L\alpha_{k})^{t}\sum_{\ell=0}^{t-1}(1+L\alpha_{k})^{-1-\ell}\notag\\ 
&\leq \exp(Lt\alpha_{k})\Eset[\|x_{k,0}^{i}- x_{\Hcal}^{\star}\|]  +  \frac{\sigma \alpha_{k}\exp(Lt\alpha_{k})}{L\alpha_{k}}\notag\\
&\leq 2\Eset[\|\xbar_{k} - x_{\Hcal}^{\star}\|] + \frac{2\sigma}{L},
\end{align*}
where the last inequality we use $L\Tcal\alpha_{k} \leq \ln(2)$ and $x_{k,0}^{i} = \xbar_{k}$. Thus, we obtain for all $t\in[0,\Tcal-1]$
\begin{align*}
&\Eset[\|x_{k,t+1}^{i} - x_{k,t}^{i}\|] \leq L\alpha_{k}\Eset[\|x_{k,t}^{i} - x_{\Hcal}^{\star}\|] + \sigma\alpha_{k}\notag\\ 
&\leq  2L\alpha_{k}\Eset[\|\xbar_{k} - x_{\Hcal}^{\star}\|] + 3\sigma\alpha_{k},
\end{align*}
which by using $x_{k,0}^{i} = \xbar_{k}$ gives  $\forall t\in[0,\Tcal-1]$ and $i\in\Hcal$
\begin{align}
\Eset[\|x_{k,t+1}^{i} - \xbar_{k}\|] &=   \Eset\Big[\big\|\sum_{\ell = 0}^{t}x_{k,\ell+1}^{i} - x_{k,\ell}^{i}\big\|\Big]\notag\\ 
&\leq    \sum_{\ell = 0}^{t}\Eset\Big[\big\|x_{k,\ell+1}^{i} - x_{k,\ell}^{i}\big\|\Big]\notag\\ 
& \leq 2L\Tcal\alpha_{k}\Eset[\|\xbar_{k}-x_{\Hcal}^{\star}\|] + 3\sigma \Tcal\alpha_{k}. \label{subsec:stochastic:T>1:xi-xbar} 
\end{align}
Similarly, we consider $i\in\Hcal$
\begin{align*}
&\Eset[\|x_{k,t+1}^{i} - x_{\Hcal}^{\star}\|^2] = \Eset[\|x_{k,t}^{i} - x_{\Hcal}^{\star} - \alpha_{k}\nabla Q^{i}(x_{k,t}^{i};X_{k,t}^{i})\|^2]\notag\\
&= \Eset[\|x_{k,t}^{i} - x_{\Hcal}^{\star}\|^2] -2\Eset[(x_{k,t}^{i} - x_{\Hcal}^{\star})^T\nabla Q^{i}(x_{k,t}^{i};X_{k,t}^{i})]\notag\\ 
&\quad + \Eset[\|\alpha_{k}\nabla Q^{i}(x_{k,t}^{i};X_{k,t}^{i})\|^2]\notag\\
&=  \Eset[\|x_{k,t}^{i} - x_{\Hcal}^{\star}\|^2] -2\Eset[(x_{k,t} - x_{\Hcal}^{\star})^T\nabla q^{i}(x_{k,t}^{i})]\notag\\ 
&\quad + \alpha_{k}^2\Eset[\|\nabla Q^{i}(x_{k,t}^{i};X_{k,t}^{i}) - \nabla q^{i}(x_{k,t}^{i})\|^2]\notag\\
&\quad + \alpha_{k}^2\Eset[\|\nabla q^{i}(x_{k,t}^{i}) - \nabla q^{i}(x_{\Hcal}^{\star})\|^2]\notag\\
&\leq (1+L^2\alpha_{k}^2)\Eset[\|x_{k,t}^{i} - x_{\Hcal}^{\star}\|^2] + \sigma^2\alpha_{k}^2,
\end{align*}
where the last inequality we use the convexity of $q^{i}$, Assumptions  \ref{asp:lipschitz} and \ref{assump:noise}. Using the relation $1+x\leq \exp(x)$ for all $x>0$ and $x_{k,0}^{i} = \xbar_{k}$, the preceding relation gives for all $t\in[0,\Tcal-1]$
\begin{align*}
&\Eset[\|x_{k,t+1}^{i} - x_{\Hcal}^{\star}\|^2]\notag\\ 
&\leq (1+L^2\alpha_{k}^2)^{t}\Eset[\|\xbar_{k} - x_{\Hcal}^{\star}\|^2] + \sigma^2\alpha_{k}^2\sum_{\ell=0}^{t-1}(1+L^2\alpha_{k}^2)^{t-\ell-1}\notag\\
&\leq \exp(L^2\Tcal\alpha_{k}^2)\Eset[\|\xbar_{k} - x_{\Hcal}^{\star}\|^2] + \frac{\sigma^2(1+L^2\alpha_{k}^2)}{L^2}\notag\\
&\leq 2\Eset[\|\xbar_{k} - x_{\Hcal}^{\star}\|^2] + \frac{2\sigma^2}{L^2},
\end{align*}
where the last inequality is due to $\exp(L^2\Tcal^2\alpha_{k}^2) \leq 2$. Using the relation above we now have for all $t\in[0,\Tcal-1]$ and $i\in\Hcal$
\begin{align}
&\Eset[\|x_{k,t+1}^{i}-\xbar_{k}\|^2] = \Eset[\|\sum_{\ell=0}^{t}\left(x_{k,\ell+1}^{i}-x_{k,\ell}^{i}\right)\|^2]\notag\\
&\leq t\sum_{\ell=0}^{t} \Eset[\|x_{k,\ell+1}^{i}-x_{k,\ell}^{i}\|^2] = t\alpha_{k}^2\sum_{\ell=0}^{t} \Eset[\|\nabla Q^{i}(x_{k,\ell}^{i};X_{k,\ell}^{i})\|^2]\notag\\
&= t\alpha_{k}^2\sum_{\ell=0}^{t} \Eset[\|\nabla Q^{i}(x_{k,\ell}^{i};X_{k,\ell}^{i}) - q^{i}(x_{k,\ell}^{i})\|^2]\notag\\
&\quad + t\alpha_{k}^2\sum_{\ell=0}^{t} \Eset[\|\nabla q^{i}(x_{k,\ell}^{i}) - \nabla q^{i}(x_{\Hcal}^{\star})\|^2]\notag\\
&\leq \Tcal^2\sigma^2\alpha_{k}^2 + L^2t\alpha_{k}^2\sum_{\ell=0}^{t}\Eset[\|x_{k,\ell}^{i} - x_{\Hcal}^{\star}\|^2]\notag\\
&\leq 2L^2\Tcal^2\alpha_{k}^2\Eset[\|\xbar_{k} - x_{\Hcal}^{\star}\|^2] + 3\Tcal^2\sigma^2\alpha_{k}^2. \label{subsec:stochastic:T>1:xi-xbar:sq} 
\end{align}
Note that we have $|\Fcal_{k}| = |\Hcal|  = N - f$. We then consider
\begin{align}
\xbar_{k+1} &= \frac{1}{|\Fcal_{k}|}\sum_{i\in\Fcal_{k}}x^{i}_{k,\Tcal}\notag\\ 
&=  \frac{1}{|\Hcal|}\left[\sum_{i\in\Hcal}x^{i}_{k,\Tcal} + \sum_{i\in\Bcal_{k}}x^{i}_{k,\Tcal} - \sum_{i\in\Hcal\setminus\Hcal_{k}}x^{i}_{k,\Tcal}\right]\notag\\  
&\stackrel{\eqref{subsec:stochastic:T>1:xi}}{=} \xbar_{k} - \frac{\alpha_{k}}{|\Hcal|}\sum_{i\in\Hcal}\sum_{t = 0}^{\Tcal-1}\nabla Q^{i}(x^{i}_{k,t};X^{i}_{k,t})\notag\\  
&\quad + \frac{1}{|\Hcal|}\left[ \sum_{i\in\Bcal_{k}}x^{i}_{k,\Tcal} - \sum_{i\in\Hcal\setminus\Hcal_{k}}x^{i}_{k,\Tcal}\right]\notag\\
&= \xbar_{k} - \Tcal\alpha_{k}\nabla q^{\Hcal}(\xbar_{k})\notag\\
&\quad - \frac{\alpha_{k}}{|\Hcal|}\sum_{i\in\Hcal}\sum_{t = 0}^{\Tcal-1}\left(Q^{i}(x^{i}_{k,t};X^{i}_{k,t}) - \nabla q^{i}(\xbar_{k})\right) \notag\\
&\quad  + \frac{1}{|\Hcal|}\Big[ \sum_{i\in\Bcal_{k}}x^{i}_{k,\Tcal} - \sum_{i\in\Hcal\setminus\Hcal_{k}}x^{i}_{k,\Tcal}\Big]\notag\\
&= \xbar_{k} - \Tcal\alpha_{k}\nabla q^{\Hcal}(\xbar_{k})\notag\\ 
&\quad - \frac{\alpha_{k}}{|\Hcal|}\sum_{i\in\Hcal}\sum_{t = 0}^{\Tcal-1}\left(\nabla Q^{i}(x^{i}_{k,t};X^{i}_{k,t}) - \nabla q^{i}(\xbar_{k})\right) \notag\\
&\quad  + \frac{1}{|\Hcal|}\Big[ \sum_{i\in\Bcal_{k}}\!(x^{i}_{k,\Tcal} - \xbar_{k}) -\!\!\! \sum_{i\in\Hcal\setminus\Hcal_{k}}\!\!(x^{i}_{k,\Tcal}-\xbar_{k})\Big],\label{subsec:stochastic:T>1:xbar}
\end{align}
where the last equality is due to $|\Bcal_{k}| = |\Hcal\setminus\Hcal_{k}|$. For convenience, we denote by 
\begin{align*}
V_{k}^{x} = \frac{1}{|\Hcal|}\sum_{i\in\Bcal_{k}}\!(x^{i}_{k,\Tcal} - \xbar_{k}) -\!\!\! \sum_{i\in\Hcal\setminus\Hcal_{k}}\!\!(x^{i}_{k,\Tcal}-\xbar_{k}).     
\end{align*}
Using \eqref{subsec:stochastic:T>1:xbar}, we consider
\begin{align}
&\left\|\xbar_{k+1} - x_{\Hcal}^{\star}\right\|^2\notag\\ 
&= \Big\|\xbar_{k} - x_{\Hcal}^{\star} - \Tcal\alpha_{k}\nabla q^{\Hcal}(\xbar_{k})\Big\|^2 + \|V_{k}^{x}\|^2\notag\\ 
&\quad + \Big\|\frac{\alpha_{k}}{|\Hcal|}\sum_{i\in\Hcal}\sum_{t = 0}^{\Tcal-1}\nabla Q^{i}(x^{i}_{k,t};X^{i}_{k,t}) - \nabla q^{i}(\xbar_{k})\Big\|^2\notag\\
&\quad - \frac{2\alpha_{k}}{|\Hcal|}\sum_{i\in\Hcal}\sum_{t = 0}^{\Tcal-1}\big(\xbar_{k} - x_{\Hcal}^{\star}\big)^T\big(\nabla Q^{i}(x^{i}_{k,t};X^{i}_{k,t}) - \nabla q^{i}(\xbar_{k})\big)\notag\\
&\quad + \frac{2\Tcal\alpha_{k}^2}{|\Hcal|} \sum_{i\in\Hcal}\sum_{t = 0}^{\Tcal-1}\nabla q^{\Hcal}(\xbar_{k})^T\big(\nabla Q^{i}(x^{i}_{k,t};X^{i}_{k,t} )- \nabla q^{i}(\xbar_{k})\big)\notag\\
&\quad + 2\big(\xbar_{k} - x_{\Hcal}^{\star} - \Tcal\alpha_{k}\nabla q^{\Hcal}(\xbar_{k})\big)^T V_{k}^{x}.\label{subsec:stochastic:T>1:xbar-x*}
\end{align}
We next analyze each term on the right-hand sides of \eqref{subsec:stochastic:T>1:xbar-x*}. First, using Assumptions \ref{asp:lipschitz} and \ref{asp:str_cvxty} we have
\begin{align}
&\Big\|\xbar_{k} - x_{\Hcal}^{\star} - \Tcal\alpha_{k}\nabla q^{\Hcal}(\xbar_{k})\Big\|^2\notag\\ 
&= \Big\|\xbar_{k} - x_{\Hcal}^{\star}\Big\|^2  -  2\Tcal\alpha_{k}\big(\xbar_{k} - x_{\Hcal}^{\star}\big)^T\nabla q^{\Hcal}(\xbar_{k})\notag\\ 
&\quad +  \Tcal^2\alpha_{k}^2\left\|\nabla q^{\Hcal}(\xbar_{k})\right\|^2\notag\\
&= \Big\|\xbar_{k} - x_{\Hcal}^{\star}\Big\|^2  -  2\Tcal\alpha_{k}\big(\xbar_{k} - x_{\Hcal}^{\star}\big)^T\nabla q^{\Hcal}(\xbar_{k})\notag\\ 
&\quad + \Tcal^2\alpha_{k}^2\left\| \nabla q^{\Hcal}(\xbar_{k})-\nabla q^{\Hcal}(x_{\Hcal}^{\star})\right\|^2\notag\\
&\leq \big(1 - 2\mu \Tcal\alpha_{k} + \Tcal^2L^2\alpha_{k}^2\big)\left\|\xbar_{k} - x_{\Hcal}^{\star}\right\|^2.\label{subsec:stochastic:T>1:Eq1a}
\end{align}
Second, by \eqref{alg:CEfilter} there exists $j\in\Hcal\setminus\Hcal_{k}$ such that $\|x^{i}_{k,\Tcal} - \xbar_{k}\| \leq \|x^{j}_{k,\Tcal} - \xbar_{k}\|$ for all $i\in\Bcal_{k}$. Then we have
\begin{align}
&\Eset[\|V_{k}^{x}\|^2]\notag\\ 
& = \Eset\Big[ \Big\|\frac{1}{|\Hcal|}\!\! \sum_{i\in\Bcal_{k}}\!\left(x^{i}_{k,\Tcal} - \xbar_{k}\right) - \frac{1}{|\Hcal|}\!\! \sum_{i\in\Hcal\setminus\Hcal_{k}}\!\!\!\left(x^{i}_{k,\Tcal}-\xbar_{k}\right)\Big\|^2\Big]\notag\\
&\leq \frac{2|\Bcal_{k}|}{|\Hcal|^2}\sum_{i\in\Bcal_{k}}\Eset[\left\|x^{i}_{k,\Tcal} - \xbar_{k}\right\|^2]\notag\\ 
&\quad + \frac{2|\Bcal_{k}|}{|\Hcal|^2}\sum_{i\in\Hcal\setminus\Hcal_{k}}\Eset[\left\|x^{i}_{k,\Tcal}-\xbar_{k}\right\|^2]\notag\\
&\leq \frac{2|\Bcal_{k}|^2}{|\Hcal|^2}\Eset[\|x^{j}_{k,\Tcal} - \xbar_{k}\|^2]\notag\\ 
&\quad + \frac{2|\Bcal_{k}|}{|\Hcal|^2}\sum_{i\in\Hcal\setminus\Hcal_{k}}\Eset[\|x^{i}_{k,\Tcal}-\xbar_{k}\|^2]\notag\\
&\stackrel{\eqref{subsec:stochastic:T>1:xi-xbar:sq}}{\leq} \frac{8\Tcal^2L^2|\Bcal_{k}|^2}{|\Hcal|^2} \alpha_{k}^2\Eset[\|\xbar_{k} - x_{\Hcal}^{\star}\|^2] + \frac{12\Tcal^2\sigma^2|\Bcal_{k}|^2}{|\Hcal|^2} \alpha_{k}^2. \label{subsec:stochastic:T>1:Eq1b}  
\end{align}
Third, using Assumptions~\ref{asp:lipschitz} and \ref{assump:noise}, and \eqref{subsec:stochastic:T>1:xi-xbar:sq}  yields
\begin{align}
&\Eset\Big[\|\frac{\alpha_{k}}{|\Hcal|}\sum_{i\in\Hcal}\sum_{t = 0}^{\Tcal-1}\left(\nabla Q^{i}(x^{i}_{k,t};X_{k,t}^{i}) - \nabla q^{i}(\xbar_{k})\right)\|^2\Big]\notag\\
&= \Eset\Big[\big\|\frac{\alpha_{k}}{|\Hcal|}\sum_{i\in\Hcal}\sum_{t = 0}^{\Tcal-1}\left(\nabla Q^{i}(x^{i}_{k,t};X_{k,t}^{i}) - \nabla q^{i}(x_{k,t}^{i})\right)\big\|^2\Big]\notag\\
&\quad + \Eset\Big[\big\|\frac{\alpha_{k}}{|\Hcal|}\sum_{i\in\Hcal}\sum_{t = 0}^{\Tcal-1}\left(\nabla q^{i}(x_{k,t}^{i}) - \nabla q^{i}(\xbar_{k})\right)\big\|^2\Big]\notag\\
&\leq \frac{\Tcal\alpha_{k}^2}{H}\sum_{i\in\Hcal}\sum_{t = 0}^{\Tcal-1} \Eset\Big[\|\left(\nabla Q^{i}(x^{i}_{k,t};X_{k,t}^{i}) - \nabla q^{i}(x_{k,t}^{i})\right)\|^2\Big]\notag\\
&\quad + \frac{\Tcal\alpha_{k}^2}{|\Hcal|}\sum_{i\in\Hcal}\sum_{t = 0}^{\Tcal-1}\Eset[\|\nabla q^{i}(x_{k,t}^{i}) - \nabla q^{i}(\xbar_{k})\|^2]\notag\\
&\leq \Tcal^2\sigma^2\alpha_{k}^2 + \frac{\Tcal L^2\alpha_{k}^2}{|\Hcal|}\sum_{i\in\Hcal}\sum_{t = 0}^{\Tcal-1}\Eset[\|x_{k,t}^{i}-\xbar_{k}\|^2]\notag\\
&\stackrel{\eqref{subsec:stochastic:T>1:xi-xbar:sq}}{\leq} 2L^4\Tcal^{4}\alpha_{k}^4\Eset[\|\xbar_{k} - x_{\Hcal}^{\star}\|^2] + 3 L^2\sigma^2\Tcal^{4}\alpha_{k}^4+\sigma^2\Tcal^2\alpha_{k}^2.\label{subsec:stochastic:T>1:Eq1c}  
\end{align}
Similarly, we consider  
\begin{align*}
&-2\Eset\Big[\big(\xbar_{k} - x_{\Hcal}^{\star}\big)^T\big(\nabla Q^{i}(x^{i}_{k,t};X^{i}_{k,t}) - \nabla q^{i}(\xbar_{k})\big)\;|\;\Pcal_{k,t}\Big]\notag\\
&=-2\big(\xbar_{k} - x_{\Hcal}^{\star}\big)^T\big(\nabla q^{i}(x^{i}_{k,t}) - \nabla q^{i}(\xbar_{k})\big)\notag\\
&\leq 2 L\|\xbar_{k} - x_{\Hcal}^{\star}\| \|x^{i}_{k,t} - \xbar_{k}\|\notag\\
&\leq L^2\Tcal\alpha_{k}\|\xbar_{k} - x_{\Hcal}^{\star}\|^2 +  \frac{1}{\Tcal\alpha_{k}}\|x^{i}_{k,t} - \xbar_{k}\|^2,
\end{align*}
where the last inequality is due to the Cauchy-Schwarz inequality $2xy \leq \eta x^2 + y^2/\eta$ for any $\eta>0$ and $x,y\in\Rset$. Taking the expectation on both sides of the preceding relation and using \eqref{subsec:stochastic:T>1:xi-xbar:sq} we obtain 
\begin{align*}
&- 2 \Eset\Big[\big(\xbar_{k} - x_{\Hcal}^{\star}\big)^{T}\big(\nabla Q^{i}(x^{i}_{k,t};X^{i}_{k,t}) - \nabla q^{i}(\xbar_{k})\big)\Big]\notag\\
&\leq L^2\Tcal\alpha_{k}\Eset[\|\xbar_{k} - x_{\Hcal}^{\star}\|^2] +  2L^2\Tcal\alpha_{k}\Eset[\|\xbar_{k} - x_{\Hcal}^{\star}\|^2]\notag\\ 
&\quad +  3\sigma^2\Tcal\alpha_{k}\notag\\
&\leq 3L^2\Tcal\alpha_{k}\Eset[\|\xbar_{k}-x_{\Hcal}^{\star}\|^2] +  3\sigma^2\Tcal\alpha_{k}.  
\end{align*}
Using the preceding relation we obtain an upper bound for the fourth term on the right-hand side of \eqref{subsec:stochastic:T>1:xbar-x*}
\begin{align}
&-\Eset\Big[\frac{2 \alpha_{k}}{|\Hcal|}\sum_{i\in\Hcal}\sum_{t = 0}^{\Tcal-1}\big(\xbar_{k} - x_{\Hcal}^{\star}\big)^T\big(\nabla Q^{i}(x^{i}_{k,t};X^{i}_{k,t}) - \nabla q^{i}(\xbar_{k})\big)\Big]\notag\\
&\leq 3L^2\Tcal^2\alpha_{k}^2\Eset[\|\xbar_{k}-x_{\Hcal}^{\star}\|^2] +  3\sigma^2\Tcal^2\alpha_{k}.
\label{subsec:stochastic:T>1:Eq1d}
\end{align}
Similarly, we provide an upper bound for the fifth term on the right-hand side of \eqref{subsec:stochastic:T>1:xbar-x*}. Indeed, using Assumption \ref{asp:lipschitz} and $\nabla q^{\Hcal}(x_{\Hcal}^{\star}) = 0$ gives
\begin{align*}
&2\Eset\Big[\nabla q^{\Hcal}(\xbar_{k})^T\big(\nabla Q^{i}(x^{i}_{k,t};X^{i}_{k,t} )- \nabla q^{i}(\xbar_{k})\big)\,|\,\Pcal_{k,t}\Big]\notag\\   
&=2\nabla q^{\Hcal}(\xbar_{k})^T\big(\nabla q^{i}(x^{i}_{k,t})- \nabla q^{i}(\xbar_{k})\big)\notag\\   
&=2\big(\nabla q^{\Hcal}(\xbar_{k})-\nabla q^{\Hcal}(x_{\Hcal}^{\star})\big)^T\big(\nabla q^{i}(x^{i}_{k,t})- \nabla q^{i}(\xbar_{k})\big)\notag\\   
&\leq 2L^2\|\xbar_{k}-x_{\Hcal}^{\star}\|\|x^{i}_{k,t} - \xbar_{k}\|\notag\\
&\leq L^3\Tcal\alpha_{k}\|\xbar_{k}-x_{\Hcal}^{\star}\|^2 + \frac{L}{\Tcal\alpha_{k}}\|x^{i}_{k,t} - \xbar_{k}\|^2,
\end{align*}
where the last inequality is due to the Cauchy-Schwarz inequality $2xy \leq \eta x^2 + y^2/\eta$ for any $\eta>0$ and $x,y\in\Rset$. Thus, by taking the expectation on both sides and using \eqref{subsec:stochastic:T>1:xi-xbar:sq} gives
\begin{align*}
&2\Eset\Big[\nabla q^{\Hcal}(\xbar_{k})^T\big(\nabla Q^{i}(x^{i}_{k,t};X^{i}_{k,t} )- \nabla q^{i}(\xbar_{k})\big)\Big]\notag\\   
&\leq L^3\Tcal\alpha_{k}\Eset[\|\xbar_{k}-x_{\Hcal}^{\star}\|^2] + 2L^3\Tcal\alpha_{k}\Eset[\|\xbar_{k} - x_{\Hcal}^{\star}\|^2]\notag\\ 
&\quad + 3L\sigma^2\Tcal\alpha_{k}\notag\\
&= 3L^3\Tcal\alpha_{k}\Eset[\|\xbar_{k}-x_{\Hcal}^{\star}\|^2] + 3L\sigma^2\Tcal\alpha_{k}.
\end{align*}
Using the above in the fifth term on the right-hand side of \eqref{subsec:stochastic:T>1:xbar-x*} yields
\begin{align}
&\Eset\Big[\frac{2\Tcal\alpha_{k}^2}{|\Hcal|} \sum_{i\in\Hcal}\sum_{t = 0}^{\Tcal-1}\nabla q^{\Hcal}(\xbar_{k})^T\big(\nabla Q^{i}(x^{i}_{k,t};X^{i}_{k,t} )- \nabla q^{i}(\xbar_{k})\big) \Big]\notag\\
&\leq  3L^{3}\Tcal^{3}\alpha_{k}^3 \Eset[\|\xbar_{k} - x_{\Hcal}^{\star}\|^2] + 3L\sigma^2\Tcal^3\alpha_{k}^3.\label{subsec:stochastic:T>1:Eq1e}
\end{align}
Finally, we analyze the last term on the right-hand side of \eqref{subsec:stochastic:T>1:xbar-x*}. Using $\nabla q^{\Hcal}(x_{\Hcal}^{\star}) = 0$, \eqref{alg:CEfilter}, \eqref{subsec:stochastic:T>1:Eq1b}, and the relation $2 \langle x, \, y \rangle \leq \eta \|x\|^2 + \|y\|^2/\eta$ for any $\eta>0$ we have
\begin{align}
&2\Eset\Big[\big(\xbar_{k} - x_{\Hcal}^{\star} - \Tcal\alpha_{k}\nabla q^{\Hcal}(\xbar_{k})\big)^T V_{k}^{x}\Big]\notag\\
&\leq \frac{3\Tcal L|\Bcal_{k}|\alpha_{k}}{|\Hcal|}\Eset[\|\xbar_{k} - x_{\Hcal}^{\star} - \Tcal\alpha_{k}\nabla q^{\Hcal}(\xbar_{k})\|^2] \notag\\
&\quad +\frac{|\Hcal|}{3\Tcal L|\Bcal_{k}|\alpha_{k}}\Eset[\|V_{k}^{x}\|^2] \notag\\
&\stackrel{\eqref{subsec:stochastic:T>1:Eq1b}}{\leq} \frac{3\Tcal L|\Bcal_{k}|\alpha_{k}}{|\Hcal|}\Eset[\|\xbar_{k} - x_{\Hcal}^{\star} - \Tcal\alpha_{k}\nabla q^{\Hcal}(\xbar_{k})\|^2] \notag\\
&\quad + \frac{8\Tcal L|\Bcal_{k}|\alpha_{k}}{3|\Hcal|} \Eset[\|\xbar_{k} - x_{\Hcal}^{\star}\|^2]  + \frac{4T\sigma^2|\Bcal_{k}|}{L|\Hcal|} \alpha_{k}\notag\\
&= \frac{17\Tcal L|\Bcal_{k}|\alpha_{k}}{3|\Hcal|}\Eset[\|\xbar_{k} - x_{\Hcal}^{\star}\|^2] + \frac{4T\sigma^2|\Bcal_{k}|}{L|\Hcal|} \alpha_{k}\notag\\
&\quad +  \frac{3\Tcal^3L|\Bcal_{k}|\alpha_{k}^3}{|\Hcal|} \Eset[\|\nabla q^{\Hcal}(\xbar_{k})\|^2]\notag\\
&\quad - \frac{6\Tcal^2L|\Bcal_{k}|\alpha_{k}^2}{|\Hcal|}\Eset[(\xbar_{k} - x_{\Hcal}^{\star})^T\nabla q^{\Hcal}(\xbar_{k})]  \notag\\
&\leq \frac{17\Tcal L|\Bcal_{k}|\alpha_{k}}{3|\Hcal|}\Eset[\|\xbar_{k} - x_{\Hcal}^{\star}\|^2] + \frac{4T\sigma^2|\Bcal_{k}|}{L|\Hcal|} \alpha_{k}\notag\\
&\quad +  \frac{3\Tcal^3L^3|\Bcal_{k}|\alpha_{k}^3}{|\Hcal|} \Eset[\|\xbar_{k} - x_{\Hcal}^{\star}\|^2]
. \label{subsec:stochastic:T>1:Eq1f}
\end{align}
Substituting from \eqref{subsec:stochastic:T>1:Eq1a}--\eqref{subsec:stochastic:T>1:Eq1f} into \eqref{subsec:stochastic:T>1:xbar-x*}, and using $L\Tcal\alpha_{k} \leq \ln(2) \leq 1$ we obtain that
\begin{align}
&\Eset[\big\|\xbar_{k+1} - x_{\Hcal}^{\star}\big\|^2\notag\\ 
&\leq  \big(1 - 2\mu \Tcal\alpha_{k}+ \Tcal^2L^2\alpha_{k}^2\big)\Eset[\big\|\xbar_{k} - x_{\Hcal}^{\star}\big\|^2]\notag\\
&\quad + \frac{8\Tcal^2L^2|\Bcal_{k}|^2}{|\Hcal|^2} \alpha_{k}^2\Eset[\|\xbar_{k} - x_{\Hcal}^{\star}\|^2] + \frac{12\Tcal^2\sigma^2|\Bcal_{k}|^2}{|\Hcal|^2} \alpha_{k}^2\notag\\
&\quad + 2L^4\Tcal^{4}\alpha_{k}^4\Eset[\|\xbar_{k} - x_{\Hcal}^{\star}\|^2] + 3 L^2\sigma^2\Tcal^{4}\alpha_{k}^4+\sigma^2\Tcal^2\alpha_{k}^2\notag\\
&\quad + 3\Tcal^2L^2\alpha_{k}^2\Eset[\|\xbar_{k}-x_{\Hcal}^{\star}\|^2] + 3\sigma^2 \Tcal^2\alpha_{k}^2\notag\\
&\quad + 3L^{3}\Tcal^{3}\alpha_{k}^3 \Eset[\|\xbar_{k} - x_{\Hcal}^{\star}\|^2] + 3L\sigma^2\Tcal^3\alpha_{k}^3\notag\\
&\quad + \frac{17\Tcal L|\Bcal_{k}|\alpha_{k}}{3|\Hcal|}\Eset[\|\xbar_{k} - x_{\Hcal}^{\star}\|^2] + \frac{4T\sigma^2|\Bcal_{k}|}{L|\Hcal|} \alpha_{k}\notag\\
&\quad +  \frac{3\Tcal^3L^3|\Bcal_{k}|\alpha_{k}^3}{|\Hcal|} \Eset[\|\xbar_{k} - x_{\Hcal}^{\star}\|^2]\allowdisplaybreaks\notag\\
&\leq \Big(1 - 2\big(\mu  - \frac{17Lf}{6|\Hcal|}\big)\Tcal\alpha_{k}  \Big)\Eset[\big\|\xbar_{k} - x_{\Hcal}^{\star}\big\|^2]\notag\\
&\quad + \frac{8\Tcal^2L^2f^2}{|\Hcal|^2} \alpha_{k}^2\Eset[\|\xbar_{k} - x_{\Hcal}^{\star}\|^2] + \frac{12\Tcal^2\sigma^2f^2}{|\Hcal|^2} \alpha_{k}^2\notag\\
&\quad + 6\Tcal^2L^2\alpha_{k}^2\Eset[\|\xbar_{k}-x_{\Hcal}^{\star}\|^2] + 7\sigma^2 \Tcal^2\alpha_{k}^2 + \frac{4T\sigma^2f}{L|\Hcal|} \alpha_{k}\notag\\
&\quad +  \frac{3\Tcal^2L^2f\alpha_{k}^2}{|\Hcal|} \Eset[\|\xbar_{k} - x_{\Hcal}^{\star}\|^2]  ,\label{subsec:stochastic:T>1:Eq1}
\end{align}
where the last inequality we use $|\Bcal_{k}|\leq f$. 
\begin{align*}
& \Eset[\big\|\xbar_{k+1} - x_{\Hcal}^{\star}\big\|^2\notag \\
&\leq \Big(1 - 2\big(\mu  - \frac{17Lf}{6|\Hcal|}\big)\Tcal\alpha_{k}  \Big)\Eset[\big\|\xbar_{k} - x_{\Hcal}^{\star}\big\|^2]\notag\\
&\quad + \frac{8\Tcal^2L^2f^2}{|\Hcal|^2} \alpha_{k}^2\Eset[\|\xbar_{k} - x_{\Hcal}^{\star}\|^2] + \frac{12\Tcal^2\sigma^2f^2}{|\Hcal|^2} \alpha_{k}^2\notag\\
&\quad + 9\Tcal^2L^2\alpha_{k}^2\Eset[\|\xbar_{k}-x_{\Hcal}^{\star}\|^2] +10\sigma^2 \Tcal^2\alpha_{k}^2 + \frac{4T\sigma^2f}{L|\Hcal|} \alpha_{k}\notag\\
&\quad +  \frac{3\Tcal^2L^2f\alpha_{k}^2}{|\Hcal|} \Eset[\|\xbar_{k} - x_{\Hcal}^{\star}\|^2].
\end{align*}

Using \eqref{thm:deterministic:T=1:f_cond} gives
\begin{align*}
\mu - \frac{17Lf}{6|\Hcal|} \geq \frac{\mu}{18}.
\end{align*}
Thus, since $f/|\Hcal| \leq \frac{1}{3}$ we obtain from \eqref{subsec:stochastic:T>1:Eq1}
\begin{align*}
&\Eset[\big\|\xbar_{k+1} - x_{\Hcal}^{\star}\big\|^2\notag\\ 
&\leq \big(1 - \frac{\mu \Tcal\alpha_{k}}{9} \big)\Eset[\big\|\xbar_{k} - x_{\Hcal}^{\star}\big\|^2]\notag\\
&\quad + \Big(6 + \frac{3f}{|\Hcal|} +  \frac{8f^2}{|\Hcal|^2}\Big)\Tcal^2L^2\alpha_{k}^2\Eset[\|\xbar_{k} - x_{\Hcal}^{\star}\|^2] \notag\\
&\quad + \frac{12\Tcal^2\sigma^2f^2}{|\Hcal|^2} \alpha_{k}^2 + 7\sigma^2 \Tcal^2\alpha_{k}^2 + \frac{3T\sigma^2f}{L|\Hcal|} \alpha_{k}\notag\\
&\leq \big(1 - \frac{\mu \Tcal\alpha_{k}}{9} \big)\Eset[\big\|\xbar_{k} - x_{\Hcal}^{\star}\big\|^2]\notag\\
&\quad + 8\Tcal^2L^2\alpha_{k}^2\Eset[\|\xbar_{k} - x_{\Hcal}^{\star}\|^2]  + 9\sigma^2 \Tcal^2\alpha_{k}^2 + \frac{3T\sigma^2f}{L|\Hcal|} \alpha_{k}.
\end{align*}
Since $\alpha_{k} = \alpha \leq \mu/(144\Tcal L^2)$ the preceding relation yields
\begin{align*}
&\Eset[\big\|\xbar_{k+1} - x_{\Hcal}^{\star}\big\|^2\notag\\ 
&\leq \Big(1 - \big(\frac{\mu}{9} - 8\Tcal L^2\alpha \big)\Tcal\alpha \Big)\Eset[\big\|\xbar_{k} - x_{\Hcal}^{\star}\big\|^2]\notag\\
&\quad + 9\sigma^2 \Tcal^2\alpha_{k}^2 + \frac{3\Tcal\sigma^2f}{L|\Hcal|} \alpha_{k}\notag\\
&\leq \Big(1-\frac{\mu \Tcal\alpha}{18}\Big)\Eset[\big\|\xbar_{k} - x_{\Hcal}^{\star}\big\|^2] + 9\sigma^2 \Tcal^2\alpha^2 + \frac{3\Tcal\sigma^2f}{L|\Hcal|} \alpha\notag\\
&\leq \Big(1-\frac{\mu \Tcal\alpha}{18}\Big)^{k}\Eset[\big\|\xbar_{0} - x_{\Hcal}^{\star}\big\|^2] + \frac{162\sigma^2 \Tcal}{\mu}\alpha + \frac{54\sigma^2f}{\mu L|\Hcal|},
\end{align*}
which conludes our proof.
\end{proof}

\end{document}